\documentclass[manuscript]{aastex}
\usepackage{txfonts}
\usepackage{graphicx}
\usepackage{aalongtable}

\slugcomment{Accepted to ApJ}

\shorttitle{Evolution of the dust/gas environment around Herbig
Ae/Be stars} \shortauthors{Liu et al.}

\begin{document}

\title{Evolution of the dust/gas environment around Herbig Ae/Be stars}

\author{Tie Liu\altaffilmark{1}, Huawei Zhang\altaffilmark{1}, Yuefang Wu\altaffilmark{1}, Sheng-Li Qin\altaffilmark{2}, Martin Miller\altaffilmark{2}}

\altaffiltext{1}{Department of Astronomy, Peking University, 100871,
Beijing China; liutiepku@gmail.com} \altaffiltext{2}{I.
Physikalisches Institut, Universit\"at zu K\"oln, Z\"ulpicher Str.
77, 50937 K\"oln, Germany}

\begin{abstract}
With the KOSMA 3-m telescope, 54 Herbig Ae/Be stars were surveyed in
CO and $^{13}$CO emission lines. The properties of the stars and
their circumstellar environments are studied by fitting the SEDs.
The mean line width of $^{13}$CO (2-1) lines of this sample is 1.87
km~s$^{-1}$. The average column density of H$_{2}$ is found to be
$4.9\times10^{21}$~cm$^{-2}$ for the stars younger than $10^{6}$~yr,
while drops to $2.5\times10^{21}$~cm$^{-2}$ for those older than
$10^{6}$~yr. No significant difference is found among the SEDs of
Herbig Ae stars and Herbig Be stars at the same age. The infrared
excess decreases with age. The envelope masses and the envelope
accretion rates decease with age after $10^{5}$~yr. The average disk
mass of the sample is $3.3\times10^{-2}~M_{\sun}$. The disk
accretion rate decreases more slowly than the envelope accretion
rate.  A strong correlation between the CO line intensity and the
envelope mass is found.

\end{abstract}


\section{Introduction}
Although significant number of astrophysical processes such as
outflow, inflow, disk, rotation have been observed towards high-mass
star formation regions,  it is still unclear whether high-mass stars
form in the same way as low-mass counterparts. Thus detail studies
of the circumstellar environment of young high-mass stars are
especially important to understand their formation and evolution
processes. However, most of high-mass stars are far away from us and
deeply embedded in the clouds, which add to the difficulties of
detailed studies. Intermediate-mass (3-8 M$_{\sun}$)
pre-main-sequence Herbig Ae/Be stars are visible at optical and
infrared wavelengths which share similar circumstellar properties
with massive stars, therefore will provide clues in understanding
the circumstellar structure of high-mass young stars. Additionally,
since Herbig Ae (HAe) stars are the precursors of vega-type systems,
investigations of their circumstellar environment are also important
for planet formation studies.

Hebig Ae/Be stars are still located in molecular clouds. More and
more observational evidences and detailed modeling of the spectral
energy distributions (SEDs) strongly suggest the presence of
circumstellar disks around Herbig Ae/Be stars
\citep{hill92,chia01,domi03}. To probe their surrounding gas
properties especially to investigate the outer part of the disk and
determine the mass and grain properties, millimeter observations are
needed \citep{albi09}. Recently, a handful of disks around Herbig
Ae/Be stars have been spatially and spectrally resolved by optical
and infrared \citep{boc03,eis04,mon08,mur10,per06,oka09}, as well as
millimeter and sub-millimeter interferometric observations
\citep{albi09,fue06,ham10,mat07,ober10,sch08,wang08}. However, the
number of disks well studied at millimeter and sub-millimeter
wavelengths is still very small, and more samples are needed. In
addition, we now have more multi-wavelength datasets, including
long-wavelength bands, than what were available for previous
studies, enabling us to model their detail structures including
star-disk-envelope.

In this paper, we report the results from a survey towards 54 Herbig
Ae/Be stars using the KOSMA 3-m telescope in CO and $^{13}$CO
emission lines. Together with SED modeling, we investigate the
properties of the circumstellar environment around these Herbig
Ae/Be stars and explore their evolution. Our studies also provide a
sample of Herbig Ae/Be stars surrounded by rich dust/gas, which are
ideal for further higher spatial resolution observations by
interferometers at millimeter and sub-millimeter wavelengths. The
next section is about sample selection, and Section 3 describes the
observation. The basic results are presented in Section 4, and more
detailed discussions are given in Section 5. In section 6, a summary
is present.

\section{Sample}
In order to collect a complete sample, we surveyed 54 Herbig Ae/Be
stars from \cite{th94} with declination $\delta>-20\arcdeg$ which
are accessible to KOSMA telescope. These sources cover a large age
range from 10$^{4}$ yr to 10$^{7}$ yr, which are very useful to
study the evolution of Herbig Ae/Be stars and their circumstellar
environments. The basic parameters of these stars are listed in
Table 1. The distances and spectral types of these stars are mainly
obtained from \cite{th94} and \cite{man06}, and completed with the
help of the SIMBAD database, operated at CDS, Strasbourg, France.
There are 27 Ae stars, 24 Be stars and 3 Fe stars in this sample.
The masses of 40 sources are obtained from \cite{man06}. The
effective temperatures are assigned from the spectral types and
listed in the seventh column of Table 1.

\section{Observations}
The single-point survey observations in $^{12}$CO (2-1), $^{12}$CO
(3-2), $^{13}$CO (2-1) and $^{13}$CO (3-2) were carried out between
January and February 2010 using the KOSMA 3-m telescope at
Gornergrat, Switzerland. The medium and variable resolution acousto
optical spectrometers with bandwidth of 300 and 655-1100 MHz at 230
and 345 GHz were used as backends. The spectral resolutions were
0.22 km/s and 0.29 km/s for 230 and 345 GHz, respectively. The beam
width at 230 and 345 GHz were 130$\arcsec$ and 82$\arcsec$,
respectively. Dual channel SIS receiver for 230/345 GHz is used for
frontend, with the typical system temperatures of 120/150 K for
230/345 GHz, respectively. The forward efficiency F$_{eff}$ was 0.93
and the corresponding main beam efficiencies B$_{eff}$ were 0.68 and
0.72 at 230 and 345 GHz, respectively. Pointing was frequently
checked on planets and was better than 10$\arcsec$. The integration
time for each source was about 8 seconds using the total power
observation mode. For the data analysis, the GILDAS software package
including CLASS and GREG was employed \citep{gui00}.

\section{Results}
\subsection{Survey results}
All of the sample sources were surveyed in $^{12}$CO (2-1) and
$^{12}$CO (3-2) lines which have been successfully detected towards
41 out of the 54 sources. However, due to the bad weather, only 28
of the 41 sources were surveyed in $^{13}$CO (2-1) and $^{13}$CO
(3-2). Fig. 1 and Fig. 2 presents the spectra of all the 54 sources.
The red and green lines represent $^{12}$CO (2-1) and $^{13}$CO
(2-1), respectively. For the sources not surveyed in $^{13}$CO
emission, only $^{12}$CO (2-1) spectra are presented.

For the 41 sources successfully detected in $^{12}$CO emission, we
fitted these observed lines with a Gaussian function and present the
fitting parameters including the $V_{LSR}$, line width (FWHM) and
antenna temperature $T_{A}^{\phantom{A}*}$ in Table 2. The
systematic velocities of each source are obtained by averaging the
$V_{LSR}$ of $^{13}$CO (2-1) and $^{13}$CO (3-2) lines if available.
For the sources not surveyed in $^{13}$CO emission, the systematic
velocities are obtained by averaging the $V_{LSR}$ of $^{12}$CO
(2-1) and $^{12}$CO (3-2) lines. The derived systematic velocities
are listed in the [col. 2] of Table 3. The spectral profiles of
these sources show rich properties. Together with $^{13}$CO (2-1)
lines, the profiles of $^{12}$CO (2-1) lines are identified and
listed in the last column of Table 2. There are 11 sources having
multiple components; 3 sources having flat tops; 5 sources showing
red asymmetry profile and 5 sources showing blue asymmetry profile.
Additionally, 2 sources are identified as blue profile, 1 source as
red profile, and 1 source shows self-absorption.

The intensity ratios of $^{12}$CO (2-1) and $^{13}$CO (2-1) are
derived and listed in the [col. 3] of Table 3. The $^{12}$CO (2-1)
optical depth can be obtained from the comparison of the observed
$^{12}$CO (2-1) and $^{13}$CO (2-1) antenna temperature, assuming
[$^{12}$CO]/[$^{13}$CO]=89. The $^{12}$CO (2-1) lines are found to
be always optically thick with an average optical depth of 26.6.
Since the $^{12}$CO is optically thick, the excitation temperature
$T_{ex}$ can be obtained according to \cite{gar91}:
\begin{equation}
T_{r}=\frac{T_{A}^{*}}{\eta_{b}}=\frac{h\nu}{k}[\frac{1}{exp(h\nu/kT_{ex})-1}\frac{1}{exp(h\nu/kT_{bg})-1}]\times~f_{\nu}
\end{equation}
where $T_{bg}=2.73~K$ is the temperature of the cosmic background
radiation,
$f_{\nu}=f_{82}=\frac{\Theta_{s}^{2}}{\Theta_{s}^{2}+82^{2}}$,
$f_{\nu}=f_{130}=\frac{\Theta_{s}^{2}}{\Theta_{s}^{2}+130^{2}}$ are
the beam-filling factors of the $^{12}$CO (3-2) and the $^{12}$CO
(2-1), respectively. The source size $\Theta_{s}$ and the excitation
temperature $T_{ex}$ can be calculated simultaneously from equ (1)
using $^{12}$CO (2-1) and $^{12}$CO (3-2) data together. To
calculate the line ratios R$_{3,2}$ of the $^{12}$CO (3-2) and the
$^{12}$CO (2-1), beam filling factor should be taken into account.
The corrected line ratios R$_{3,2}$ can be obtained from:
$R_{3,2}=f_{82,130}\cdot\frac{\int~T_{A}(12CO(3-2))dv}{\int~T_{A}(12CO(2-1))dv}$,
where
$f_{82,130}=\frac{1+(82/\Theta_{s})^{2}}{1+(130/\Theta_{s})^{2}}$
and $\Theta_{s}$ is the source size.

Assuming the $^{13}$CO (2-1) is optically thin, the optical depth
and column density of the $^{13}$CO can be straightforwardly
obtained under local thermal equilibrium (LTE) assumption. With the
abundance ratio [H$_{2}$]/[$^{13}$CO]=$8.9\times10^{5}$, the column
density of H$_{2}$ also can be calculated. For these sources which
have not been surveyed with the $^{13}$CO, the column densities of
H$_{2}$ are firstly obtained from the $^{12}$CO (2-1) lines under
the assumption that the $^{12}$CO (2-1) emission is optically thin
and [H$_{2}$]/[$^{12}$CO]=$10^{4}$, and then multiplied by an
optical depth correction factor $C_{\tau}=\tau/(1-e^{-\tau})$. The
optical depths of the $^{12}$CO (2-1) for those sources not surveyed
in the $^{13}$CO emission are assigned as 26.6, the average value
obtained above. All the results are collected in Table 3.

\subsection{The SED}
We have compiled the SEDs for 53 sources in the sample except MWC
614, which has not enough data to use for modeling. The optical
(UBV) data are from \cite{man06}, JHK fluxes from the 2MASS All-Sky
Catalog of Point Sources, IRAC and MIPS data from
\cite{har07,gute09}, IRAS data from the IRAS catalogue of Point
Sources, MSX data from MSX6C Infrared Point Source Catalog, AKAIR
data from AKARI/IRC mid-IR all-sky Survey and AKARI/FIS All-Sky
Survey Point Source Catalogues, SCUBA 450 $\micron$ and 850
$\micron$ submillimeter data from \cite{di08}, 1.3 mm continuum data
from \cite{hen94}. We also collected sub-mm and mm data for some
sources from \cite{man94}, \cite{acke04}, \cite{di97}, and
\cite{albi09}. Although not all the sources have data at all the
wavelength bands, there are still enough data for each source to be
used for fitting the SED.

The SEDs are then modeled using the online 2D-radiative transfer
tool developed by \cite{ro06}, which has been successfully tested by
\cite{ro07}, on a sample of low-mass YSOs and by \cite{gra09} on
high-mass protostars. The geometry in the models consists of a
central star surrounded by a flared accretion disk, a rotationally
flattened infalling envelope and bipolar cavities. The SED fitting
tool uses 200,000 YSO model SEDs (20,000 sets of physical parameters
and 10 viewing angles). The distance and visual extinction are the
only two input parameters in the fitting process. During the fitting
process, all the model SEDs in the grid are compared with the
observed data, and the model SEDs that fit a source well can be
picked out within a specified limit on $\chi^{2}$. For each source,
a minimum of three data points that are of good quality are required
by the SED model tool. In our sample, all the 53 sources have at
least ten such data points, spreading over the wavelength region of
0.3 to 2600 $\micron$, which can well constraint the models with
least standard deviations \citep{ro07}. However, the two input
parameters of the distance and visual extinction are far from fixed,
which lead to the degeneracy of the models. In our cases, we assumed
30\% uncertainty for each of the distances obtained from
literatures, and varied the visual extinction A$_{v}$ from 0 to 30.
Although the best fit model has the least $\chi^{2}$ value, it may
not actually represent the data very well \citep{gra09}. In order to
find a representative set of values for each source, all the models
are accepted and analyzed, which satisfy the criterion of
$\chi^{2}-\chi_{best}^{2}<3\times n_{data}$, where $n_{data}$ is the
number of the data points. Fig. 3 and Fig. 4 shows the observed SEDs
and model SEDs with $\chi^{2}-\chi_{best}^{2}<3\times n_{data}$.
Following \cite{gra09}, a weighted mean and standard deviation are
derived for all the parameters of each source, with the weights
being the inverse of the $\chi^{2}$ of each model. All the
parameters derived are listed in Table 4. The various columns in
Table 4. are as follows: [col. 1] name of the source;[col. 2] visual
extinction; [col. 3] age of the star; [col. 4] mass of the star;
[col. 5] radius of the star; [col. 6] total luminosity of the star;
[col. 7] temperature of the star; [col. 8] envelope mass; [col. 9]
envelope accretion rate; [col. 10] disk mass ; [col. 11] disk
inclination angle; [col. 12] outer radius of the disk; [col. 13]
disk accretion rate.

\section{Discussion}
\subsection{The CO emission}
The mean line width of the $^{13}$CO(2-1) lines of this sample is
1.87 km~s$^{-1}$ with a standard deviation of 0.66 km~s$^{-1}$. This
line width is similar to that of intermediate-mass star formation
regions ($\sim$2 km~s$^{-1}$) \citep{sun06}, larger than that of
low-mass star formation regions (1.3 km~s$^{-1}$) \citep{my83}, but
much smaller than that of high-mass star formation regions
associated with IRAS sources (3.09 km~s$^{-1}$) \citep{wang09,wu03}.

The average column density of H$_{2}$ is found to be
$4.9\times10^{21}$~cm$^{-2}$ before $10^{6}$~yr, while drops to
$2.5\times10^{21}$~cm$^{-2}$ after $10^{6}$~yr. As shown in the left
panel of Fig.5, the optical depth of the $^{12}$CO(2-1) ranges from
7.5 to 54 with an average value of 27.7 before $10^{6}$~yr, but the
optical depth drops below 26.6 (the average value in this sample)
after $10^{6}$~yr, with an average value of 22.5. The excited
temperature of CO ranges from 4.3 K to 33 K, with an average of 13.8
K. As shown in the right panel of Fig.5, a decreased trend is found
for the excited temperature with age. The decrease of the optical
depth and excitation temperature with age may be due to the
different gas dispersion mechanisms in which most of gas should be
dispersed by outflows and stellar winds at the early evolutionary
stage of the stars, and UV radiation and photodissociation play an
important role in further gas dispersion as the stars evolve
\citep{fue02}. Therefore only less dense molecular gas with low
excitation temperature existed in outer envelope of late-type stars
is observed at millimeter and submillimeter wavelengths.

\subsection{The properties of the stars}
The ages of the sample stars spread from 10$^{3}$~yr to 10$^{7}$~yr.
About 40\% sample stars are younger than 10$^{6}$~yr, and the others
are older than 10$^{6}$~yr. The average mass of the stars is
$3.9\pm2.2$~M$_{\sun}$. In the left panel of Fig.6, we compare the
masses derived from SED fitting to those obtained from \cite{man06}.
The masses (M$_{mod}$) from the SED modeling are well coincident
with the masses (M$_{lit}$) obtained from literatures with a
relationship of M$_{mod}$=M$_{lit}\pm$0.44, where the correlation
coefficient r=0.64. Although the masses derived from the SED fitting
seems slightly higher than those from literatures, they are found to
agree with each other better than $\pm0.3$ orders of magnitude. We
also compare the effective temperatures of the stars obtained from
SED fitting to those corresponding to the spectral types. We find
the agreement of the effective temperatures obtained from the two
independent methods for 80\% of the sample sources is better than
$\pm0.3$ orders of magnitude. The average effective temperature of
the sample stars is (1.1$\pm$0.6)$\times10^{4}$ K. As shown in
Fig.7, the mass-luminosity function of the sample stars is found to
be
$Log(L_{*}/L_{\sun})=(3.08\pm0.17)Log(M_{*}/M_{\sun})+(0.48\pm0.10)$,
with the correlation coefficient r=0.93.

\subsection{The evolution of the SED}
\cite{mee01} proposed that the SEDs of Herbig Ae/Be stars can be
divided into two groups based on the different mid-infrared
excesses: Group I sources with strong mid-infrared flux while Group
II sources with weaker infrared excesses. \cite{arke05} found most
HAe stars belong to Group I and most HBe stars belong to Group II.
However, there may be an evolutionary link between Group I and Group
II sources. We display the SEDs of the HAe stars and HBe stars in
our sample in Fig.3 and Fig.4 ordered by age (age increases from
top-left to bottom-right panel), respectively. Firstly, we find no
significant difference among the SEDs of HAe stars and HBe stars at
the same age; secondly, it is clearly to see the infrared excess
decreases with age for both HAe stars and HBe stars. The dashed
lines show the SEDs of the stellar photosphere, including the effect
of foreground extinction, from the best-fitting model \citep{ro06}.
As shown in the second column of Table 4, the interstellar
extinctions A$_{v}$ range from 0 to 12 mag in our sample from the
SED modeling. We find at early ages ($<10^{6}$~yr), the optical and
near-infrared fluxes of most of the HAe and HBe stars can not be
simply fitted by a stellar photosphere with interstellar extinction,
while the older ones ($>10^{6}$~yr) can be well fitted, indicating
huge amounts of circumstellar dust/gas exist at early ages, which
causes large extinction and infrared excess. The shortest wavelength
at which infrared excess is recognizable is around 2 $\micron$. At
early ages, the excess at sub-mm/mm regions is also very significant
and HAe stars seem to have stronger sub-mm/mm excess than HBe stars
at the same age.

\subsection{The circumstellar envelopes and the disks}
The left panel of Fig.8 gives the number distribution of the
envelope mass and presents the envelope mass as function of stellar
age. There is no difference in the distribution of the envelope
masses between HAe stars and HBe stars. The envelope masses of the
HAe and HBe stars are larger than $10^{-2}~M_{\sun}$ before
$10^{6}$~yr, but smaller than $10^{-2}~M_{\sun}$ after $10^{6}$~yr.
The envelope masses decease with age after $10^{5}$~yr following the
relationship:$Log(M_{env}/M_{\sun})=(-4.0\pm0.5)Log(t/yr)+(22\pm3.4)$,
with the correlation coefficient r=0.74. The number distribution of
the envelope accretion rate and the envelope accretion rate as
function of stellar age are presented in the right panel of Fig.8.
The average envelope accretion rate for the Herbig Ae/Be stars
younger than $10^{6}$~yr is
$2.5\times10^{-5}$~M$_{\sun}\cdot$yr$^{-1}$. The envelope accretion
rate drops below $10^{-6}$~M$_{\sun}\cdot$yr$^{-1}$ after
$10^{6}$~yr, and 25 sources older than $10^{6}$~yr have zero
envelope accretion rate, which means those sources can be explained
by disk-only sources without infalling envelope. We also find the
envelope accretion rate decease with age after $10^{5}$~yr following
the
relationship:$Log(\dot{M}_{env}/M_{\sun}~\cdot~yr^{-1})=(-2.4\pm0.5)Log(t/yr)+(8.2\pm3.0)$,
with the correlation coefficient r=0.72.

From the left panel of Fig.9, it can be seen the disk masses locate
at a narrow range from $10^{-3}~M_{\sun}$ to $10^{-1}~M_{\sun}$ if
neglecting the three points with extremely small disk mass. The
average disk mass is $3.3\times10^{-2}~M_{\sun}$, and the maximum
one is $0.26~M_{\sun}$ from AS 447. No obvious evolutionary pattern
is found towards the disk masses. The right panel of Fig. 9 shows
the disk accretion rate as function of stellar age. The average disk
accretion rate is $5\times10^{-8}$~M$_{\sun}\cdot$yr$^{-1}$. An
average disk accretion rate of
$3.6\times10^{-7}$~M$_{\sun}\cdot$yr$^{-1}$ is found for those
sources younger than $10^{6}$~yr, which is much lower than the
corresponding envelope accretion rate, suggesting the envelopes
dominate the accretion at early stages of star formation. However,
after $10^{6}$~yr, the envelope accretion halts
($\dot{M}_{env}\sim0$), while the disk accretion rate remains
$\sim1.6\times10^{-8}$~M$_{\sun}\cdot$yr$^{-1}$. Additionally, the
disk accretion rate deceases with age following the
relationship:$Log(\dot{M}_{disk}/M_{\sun}~\cdot~yr^{-1})=(-0.6\pm0.1)Log(t/yr)+(3.5\pm0.7)$,
with the correlation coefficient r=0.6, nevertheless the disk
accretion rate deceases much more slowly than the envelope accretion
rate.

\subsection{The relationship between CO gas and the envelope}
Fig.10 presents the line intensity of the $^{12}$CO (2-1) and
$^{12}$CO (3-2) as function of stellar age. It can be seen the CO
line intensities of most of the sources younger than $10^{6}$~yr are
larger than 10 K~km~s$^{-1}$, and drop below 10 K~km~s$^{-1}$ after
$10^{6}$~yr. The evolution behaviors of CO line intensity is very
similar to that of envelope (see Fig.8), suggesting possible
co-evolution between the envelope and the CO gas. We plot the line
intensity of the $^{12}$CO (2-1) and the $^{12}$CO (3-2) as function
of envelope mass in Fig.11. Exclude those data with line intensities
smaller than 1 K~km~s$^{-1}$, which have low signal-to-noise level
or no detection of CO emission, the CO line intensity is well
correlated with the envelope mass. For $^{12}$CO (2-1), the
relationship can be fitted by a power-law:$Log(I_{CO
(2-1)}/K~km~s^{-1})=(0.076\pm0.017)Log(M_{env}/M_{\sun})+(1.129\pm0.052)$,
where the correlation coefficient r=0.64. For $^{12}$CO(3-2), the
relationship can also be fitted by a power-law:$Log(I_{CO
(3-2)}/K~km~s^{-1})=(0.098\pm0.019)Log(M_{env}/M_{\sun})+(1.037\pm0.066)$,
with a correlation coefficient r=0.64. It seems that CO gas is
associated with the envelopes.

\section{Summary}
We have surveyed 54 Herbig Ae/Be stars using the KOSMA 3-m telescope
in the CO and the $^{13}$CO emission lines. Together with SED
modeling, we have investigated the properties of the circumstellar
environments around those Herbig Ae/Be stars and explored their
evolutionary characteristics. The main findings of this paper are as
below:

(1).We have successfully detected CO emission towards 41 out of the
54 sources, and 28 of the 41 sources were surveyed in $^{13}$CO
(2-1) and $^{13}$ CO(3-2). The mean line width of the $^{13}$CO
(2-1) line width of this sample is 1.87 km~s$^{-1}$, which is
similar to that of intermediate-mass star formation regions ($\sim$2
km~s$^{-1}$). The average column density of H$_{2}$ is found to be
$4.9\times10^{21}$~cm$^{-2}$ before $10^{6}$~yr, while drops to
$2.5\times10^{21}$~cm$^{-2}$ after $10^{6}$~yr.

(2).The SEDs of 53 sources are compiled and modeled using the online
2D-radiative transfer tool. From the SED fitting, we find 40\%
sample stars are younger than 10$^{6}$~yr, and the others are older
than 10$^{6}$~yr. The average mass of the stars is
$3.9\pm2.2$~M$_{\sun}$. The mass-luminosity function of the sample
stars can be described as
$Log(L_{*}/L_{\sun})=(3.08\pm0.17)Log(M_{*}/M_{\sun})+(0.48\pm0.10)$.

(3).We find no significant difference among the SEDs of HAe stars
and HBe stars at the same age and the infrared excess decreases with
age for both HAe stars and HBe stars.

(4).The envelop masses decease with age after $10^{5}$~yr. The
average envelop accretion rate for the Herbig Ae/Be stars younger
than $10^{6}$~yr is $2.5\times10^{-5}$~M$_{\sun}\cdot$yr$^{-1}$ and
drops below $10^{-6}$~M$_{\sun}\cdot$yr$^{-1}$ after $10^{6}$~yr. In
our sample 25 sources older than $10^{6}$~yr have zero envelop
accretion rate. The average disk mass of the sample is
$3.3\times10^{-2}~M_{\sun}$. The disk accretion rate deceases with
age, but more slowly than the envelope accretion rate.

(5).The CO line intensities of most of the sources younger than
$10^{6}$~yr are larger than 10 K~km~s$^{-1}$, and drop below 10
K~km~s$^{-1}$ after $10^{6}$~yr.  The CO line intensity is well
correlated with the envelope mass, suggesting possible co-evolution
between the envelope and the CO gas.

\section*{Acknowledgment}
\begin{acknowledgements}
This work was supported by the NSFC under grants No. 11073003,
10733030, and 10873019, and by the National key Basic Research
Program (NKBRP) No. 2007CB815403.
\end{acknowledgements}

\clearpage
\begin{deluxetable}{ccrrrrrrrrrcrl}
\tabletypesize{\scriptsize}  \tablecaption{Parameters of all the
Herbig Ae/Be stars surveyed}
 \tablewidth{0pt} \tablehead{
  Name& $\alpha$(J2000) & $\delta$(J2000) &  Distance  & Spectral Type  & M$_{*}$ & log(T$_{eff}$) \\
      & (h m s)& ($\arcdeg \arcmin \arcsec$) & (pc)& & M$_{\sun}$
      &log(K)
   }
\startdata
MacC H12     &        00 07 02.6   &    65 38 38.2   &     850    &     A5 D         &           & 3.91             \\
LkHA 198 / V633cas\tablenotemark{b}     & 00 11 26.0   &    58 49 29.1   &  600    &     B9 D         &4.25       & 4.02       \\
Vx Cas       &        00 31 30.7   &    61 58 51.0   &     760    &     A0 D         &3          & 3.98         \\
RNO 6        &        02 16 30.1   &    55 22 57.0   &     1600   &     B3 D         &5.99       & 4.27             \\
IP Per       &        03 40 47.0   &    32 31 53.7   &     350    &     A6 D         &2          & 3.92         \\
XY Per       &        03 49 36.2   &    38 58 55.5   &     160    &     A2IIv+ C     &2          & 3.91       \\
V892 Tau\tablenotemark{a,b}     &        04 18 40.6   &    28 19 15.5   &     160    &     B8 D         &$>$5.11    & 4.08          \\
AB Aur\tablenotemark{a,b}      &        04 55 45.8   &    30 33 04.3   &     144    &     A0Vpe C      &2.77       & 3.97       \\
MWC 480\tablenotemark{a,b}       &        04 58 46.3   &    29 50 37.0   &     131    &     A3Ve D       &1.99       & 3.94         \\
HD 35929     &        05 27 42.8   &   -08 19 38.4   &     345    &     F2III D      &3.41       & 3.84         \\
HD 36112     &        05 30 27.5   &    25 19 57.1   &     205    &     A5 IVe       &2.17       & 3.91         \\
HD 245185    &        05 35 09.6   &    10 01 51.5   &     400    &     A1 D         &2.07       & 3.97         \\
T Ori\tablenotemark{a}        &        05 35 50.4   &   -05 28 34.9   &     460    &     A3V C        &3.34       & 3.98         \\
CQ Tau\tablenotemark{a,b}       &        05 35 58.5   &    24 44 54.1   &     100    &     F3 D         &1.5        & 3.83              \\
V380 Ori     &        05 36 25.4   &   -06 42 57.7   &     510    &     A1e D        &$>$4.93    & 3.97           \\
V586 Ori     &        05 36 59.3   &   -06 09 16.4   &     510    &     A2V C        &3          & 3.98             \\
BF Ori       &        05 37 13.3   &   -06 35 00.6   &     430    &     A5II-IIIev C &2.5        & 3.95         \\
HD37411      &        05 38 14.5   &   -05 25 13.3   &            &     B9Ve         &           & 4.02             \\
Haro 13A     &        05 38 18.2   &   -07 02 25.9   &     460    &     Be           &           &              \\
V599 Ori     &        05 38 58.6   &   -07 16 45.6   &     360    &     A5 D         &           & 3.91             \\
RR Tau       &        05 39 30.5   &   26 22 27.0    &     800    &     A2II-IIIe C  &4.26       & 3.98         \\
V350 Ori     &        05 40 11.8   &  -09 42 11.1    &     460    &     A1 D         &2.22       & 3.97         \\
MWC 789      &        06 01 60.0   &   16 30 56.7    &     700    &     B9 D         &4.13       & 4.02             \\
LkHA 208     &        06 07 49.5   &   18 39 26.5    &     1000   &     A7 D         &3.24       & 3.89         \\
LkHA 339     &        06 10 57.8   &  -06 14 37      &     830    &     A1 D         &3.18       & 3.97         \\
LkHA 215\tablenotemark{b}     &        06 32 41.8   &   10 09 33.6    &     800    &     B6 D         &$>$5.43    & 4.15           \\
R Mon\tablenotemark{a,b}        &        06 39 10     &   08 44 09.7    &     800    &     B0           &$>$5.11    & 4.08           \\
V590 Mon     &        06 40 44.6   &   09 48 02.1    &    800     &     B7 D         &$<$3.35    & 4.11           \\
GU CMa       &        07 01 49.5   &  -11 18 03.3    &            &     B2Vne C      &           & 4.34             \\
HD 141569\tablenotemark{a}     &        15 49 57.7   &  -03 55 16.3    &     99     &     A0Ve D       &2.18       & 3.98         \\
VV Ser\tablenotemark{a,b}      &        18 28 47.9   &   00 08 39.8    &     330    &     B6 D         &$>$5.43    & 4.15          \\
MWC 300      &        18 29 25.7   &  -06 04 37.1    &     650    &     Bpe D        &           &              \\
AS 310       &        18 33 27     &  -04 58 06      &     2500   &     B1           &$>$6       & 4.4            \\
MWC 614      &        19 11 04.3   &   21 11 26      &            &                  &           &              \\
Par 21       &        19 29 00.8   &   09 38 46.7    &     300    &     A5e          &           & 3.91          \\
V1295 Aql\tablenotemark{a}    &        20 03 02.5   &   05 44 16.7    &     290    &     A2IVe D      &3.03       & 3.95             \\
V1685 Cyg\tablenotemark{a}   &        20 20 28.2   &   41 21 51.6    &     980    &     B3 D         &$>$5.99    & 4.27               \\
Par 22       &        20 24 29.5   &   42 14 03.7    &            &     A5 eV        &           & 3.91             \\
PV Cep\tablenotemark{b}      &        20 45 53.9   &   67 57 38.9    &     500    &     A5 D         &           & 3.91             \\
AS 442\tablenotemark{a}       &        20 47 37.5   &   43 47 25.0    &     826    &     B9 D         &           & 4.02         \\
LkHA 134     &        20 48 04.8   &   43 47 25.8    &     700    &     B2 D         &$>$6       & 4.34           \\
HD 200775\tablenotemark{a}    &        21 01 36.9   &   68 09 47.8    &     429    &     B2 Ve C      &$>$5.99    & 4.27        \\
LkHA 324     &        21 03 54.2   &   50 15 10.2    &     780    &     B8 D         &$>$5.11    & 4.08               \\
HD 203024    &        21 16 03     &   68 54 52.1    &            &     A D          &           &          \\
V645 Cyg     &        21 39 58.2   &   50 14 21.2    &     3500   &     A0 D         &$>$4.94    & 3.98           \\
LkHA 234     &        21 43 02.3   &   66 06 29      &     1250   &     B5 Vev       &$>$5.29    & 4.11           \\
AS 477 / BD 46 &        21 52 34.1   &   47 13 43.6    &     1200   &     B9.5Ve C   &$>$4.94    & 3.98         \\
LkHA 257     &        21 54 18.8   &   47 12 09.7    &     900    &     B5 D         &           & 4.19             \\
BH Cep       &        22 01 42.9   &   69 44 36.5    &     450    &     F5IV C       &1.73       & 3.81         \\
SV Cep       &        22 21 33.2   &   73 40 27.1    &     440    &     A0 D         &2.5        & 3.98             \\
V375 Lac     &        22 34 41     &   40 40 04.5    &     880    &     A4 D         &3.19       & 3.93             \\
IL Cep       &        22 53 15.6   &   62 08 45      &     725    &     B2IV-Vne C   &           & 4.34         \\
MWC 1080\tablenotemark{a,b}     &        23 17 25.6   &   60 50 43.6    &     2200   &     B0eq D       &$>$6       & 4.48              \\
LkHA 259     &        23 58 41.6   &   66 26 12.6    &     890    &     A9 D         &           & 3.87         \\
\enddata
\tablenotetext{a}{Circumstellar disks detected by optical or
infrared observations \citep{boc03,eis04,mon08,mur10,per06,oka09}.}
\tablenotetext{b}{Circumstellar disks detected by millimeter and
sub-millimeter interferometric observations
\citep{albi09,fue06,ham10,mat07,ober10,sch08,wang08}.}
\end{deluxetable}

\clearpage
\begin{deluxetable}{ccrrrrrrrrrrrrrrrrrrcrl}
\tabletypesize{\scriptsize} \setlength{\tabcolsep}{0.05in} \rotate
\tablecaption{Observational Parameters of the lines}
 \tablewidth{0pt} \tablecolumns{17}\tablehead{
\multicolumn{2}{c}{Name} & \multicolumn{3}{c}{$^{13}$CO (2-1)}
&\colhead{}& \multicolumn{3}{c}{$^{13}$CO (3-2)} &\colhead{}&
\multicolumn{3}{c}{CO (2-1)}&\colhead{} & \multicolumn{3}{c}{CO (3-2)} & \multicolumn{2}{c}{Notes}\\
\cline{3-5} \cline{7-9} \cline{11-13} \cline{15-17}\\
\multicolumn{2}{c}{} & \colhead{V$_{LSR}$}   & \colhead{FWHM}    &
\colhead{T$_{A}^{*}$}&\colhead{} & \colhead{V$_{LSR}$}   &
\colhead{FWHM}    & \colhead{T$_{A}^{*}$} &\colhead{}&
\colhead{V$_{LSR}$}   & \colhead{FWHM}    & \colhead{T$_{A}^{*}$}
&\colhead{}& \colhead{V$_{LSR}$}   & \colhead{FWHM}    &
\colhead{T$_{A}^{*}$}&\multicolumn{2}{c}{}\\
\multicolumn{2}{c}{} & \colhead{(km~s$^{-1}$)}   &
\colhead{(km~s$^{-1}$)} & \colhead{(K)}&\colhead{} &
\colhead{(km~s$^{-1}$)}   & \colhead{(km~s$^{-1}$)}    &
\colhead{(K)} &\colhead{}& \colhead{(km~s$^{-1}$)}   &
\colhead{(km~s$^{-1}$)}    & \colhead{(K)} &\colhead{}&
\colhead{(km~s$^{-1}$)}   & \colhead{(km~s$^{-1}$)}    &
\colhead{(K)}&\multicolumn{2}{c}{} } \startdata
\multicolumn{2}{c}{MacC H12  } &     $ -4.82\pm0.02$ & $1.67\pm0.04$ &  2.64  & & $ -4.63\pm0.07$ & $1.03\pm0.16$ & 1.57  & &   $  -5.05\pm0.02 $ & $2.30\pm0.03$ &    6.98  & & $ -5.14\pm0.04$ &$2.50\pm0.09$ & 7.92   & \multicolumn{2}{c}{ flat top           }\\
\multicolumn{2}{c}{LkHA 198  } &     $ -0.25\pm0.01$ & $2.25\pm0.02$ &  3.00  & & $ -0.06\pm0.03$ & $1.98\pm0.06$ & 2.59  & &   $  -0.14\pm0.01 $ & $3.07\pm0.01$ &    6.84  & & $ -0.06\pm0.02$ &$2.67\pm0.05$ & 6.06   & \multicolumn{2}{c}{ wings              }\\
\multicolumn{2}{c}{RNO 6     } &     $-36.20\pm0.02$ & $1.59\pm0.03$ &  1.50  & & $-35.87\pm0.04$ & $1.51\pm0.10$ & 1.41  & &   $ -36.31\pm0.02 $ & $2.34\pm0.04$ &    2.81  & & $-36.27\pm0.09$ &$2.31\pm0.21$ & 2.50   & \multicolumn{2}{c}{ flat top           }\\
\multicolumn{2}{c}{XY Per    } &     $ -4.20\pm0.07$ & $2.11\pm0.18$ &  0.47  & &                 &               &       & &   $  -4.87\pm0.04 $ & $3.28\pm0.09$ &    2.23  & & $ -4.73\pm0.10$ &$2.68\pm0.30$ & 2.47   & \multicolumn{2}{c}{ red asy            }\\
\multicolumn{2}{c}{V892 Tau  } &     $  7.40\pm0.01$ & $1.33\pm0.02$ &  2.14  & & $  6.98\pm0.05$ & $1.49\pm0.15$ & 1.12  & &   $   7.09\pm0.01 $ & $2.92\pm0.03$ &    4.14  & & $  6.89\pm0.02$ &$2.31\pm0.05$ & 3.64   & \multicolumn{2}{c}{ blue asy           }\\
\multicolumn{2}{c}{AB Aur    } &                     &               &        & &                 &               &       & &   $   6.10\pm0.01 $ & $1.04\pm0.01$ &    4.70  & & $  6.18\pm0.01$ &$0.96\pm0.02$ & 4.96   & \multicolumn{2}{c}{                    }\\
\multicolumn{2}{c}{T Ori     } &     $  7.47\pm0.01$ & $1.28\pm0.03$ &  4.27  & & $  7.48\pm0.03$ & $1.19\pm0.06$ &  8.58 & &   $   7.29\pm0.21 $ & $1.78\pm0.21$ &    13.51 & & $ 7.20\pm0.01$  &$1.58\pm0.01$ & 24.73  & \multicolumn{2}{c}{ three comp         }\\
\multicolumn{2}{c}{          } &     $ 10.83\pm0.12$ & $2.50\pm0.18$ &  2.76  & & $ 11.15\pm0.25$ & $1.99\pm0.55$ &  3.07 & &   $  10.42\pm0.21 $ & $3.07\pm0.21$ &    14.38 & & $ 10.34$ &$2.73\pm0.03$ & 19.27  & \multicolumn{2}{c}{                    }\\
\multicolumn{2}{c}{          } &     $ 13.13\pm0.16$ & $2.32\pm0.25$ &  1.86  & & $ 13.34\pm0.62$ & $2.02\pm0.84$ &  1.35 & &   $  13.30\pm0.21 $ & $3.37\pm0.21$ &    11.40 & & $ 13.17$ &$3.42\pm0.04$ & 14.13  & \multicolumn{2}{c}{                    }\\
\multicolumn{2}{c}{V380 Ori  } &     $  6.99\pm0.05$ & $1.35\pm0.11$ &  2.34  & & $  6.93\pm0.09$ & $0.82\pm0.17$ & 1.85  & &                     &               &          & &                                &        &                                         \\
\multicolumn{2}{c}{          } &     $  8.95\pm0.03$ & $2.14\pm0.06$ &  6.44  & & $  9.02\pm0.05$ & $2.03\pm0.12$ & 5.60  & &   $   8.85 $        & $4.01\pm0.02$ &    7.46  & & $  8.95\pm0.02$ &$3.40\pm0.05$ & 6.50   & \multicolumn{2}{c}{ two comp           }\\
\multicolumn{2}{c}{V586 Ori  } &                     &             &          & &                 &               &       & &   $   6.54\pm0.03 $ & $1.77\pm0.07$ &    2.18  & & $  6.54\pm0.06$ &$1.93\pm0.21$ & 1.70   & \multicolumn{2}{c}{ two comp           }\\
\multicolumn{2}{c}{          } &                     &             &          & &                 &               &       & &   $   8.74\pm0.01 $ & $1.42\pm0.02$ &    7.43  & & $  8.75\pm0.01$ &$1.31\pm0.03$ & 9.09   & \multicolumn{2}{c}{                    }\\
\multicolumn{2}{c}{BF Ori    } &     $  6.59$        & $1.64\pm0.05$ &  2.86  & & $  5.77\pm0.07$ & $1.12\pm0.16$ & 2.46  & &   $   6.54\pm0.21 $ & $2.82\pm0.21$ &    6.75  & & $  6.05\pm0.04$ &$2.46\pm0.06$ & 6.53   & \multicolumn{2}{c}{ three comp         }\\
\multicolumn{2}{c}{          } &     $  9.17\pm0.09$ & $1.82\pm0.18$ &  1.49  & &                 &               &       & &   $   9.14\pm0.21 $ & $2.32\pm0.21$ &    3.98  & & $  8.77\pm0.05$ &$2.75\pm0.13$ & 4.36   & \multicolumn{2}{c}{                    }\\
\multicolumn{2}{c}{          } &     $ 10.89\pm0.10$ & $1.22\pm0.24$ &  0.88  & &                 &               &       & &   $  10.70\pm0.21 $ & $1.56\pm0.21$ &    1.79  & & $ 10.40\pm0.02$ &$1.22\pm0.08$ & 1.32   & \multicolumn{2}{c}{                    }\\
\multicolumn{2}{c}{Haro 13A  } &     $  5.78\pm0.02$ & $2.10\pm0.05$ &  2.38  & & $  5.47\pm0.07$ & $1.01\pm0.19$ & 2.17  & &   $   5.71\pm0.02 $ & $3.62\pm0.02$ &    6.00  & & $  5.19\pm0.08$ &$2.92\pm0.07$ & 5.14   & \multicolumn{2}{c}{ blue asy           }\\
\multicolumn{2}{c}{V599 Ori  } &     $  5.02\pm0.04$ & $2.40\pm0.12$ &  1.35  & & $  5.07\pm0.20$ & $2.73\pm0.53$ & 0.67  & &   $   5.25\pm0.02 $ & $3.51\pm0.03$ &    4.78  & & $  4.94\pm0.15$ &$3.49\pm0.11$ & 3.24   & \multicolumn{2}{c}{ two comp?          }\\
\multicolumn{2}{c}{          } &     $  7.18\pm0.05$ & $1.24\pm0.13$ &  0.77  & &                 &               &       & &                     &               &          & &                 &              &        & \multicolumn{2}{c}{                    }\\
\multicolumn{2}{c}{RR Tau    } &     $ -5.40\pm0.02$ & $1.36\pm0.06$ &  1.55  & & $ -5.41\pm0.04$ & $0.98\pm0.10$ & 1.14  & &   $  -5.09\pm0.01 $ & $1.83\pm0.02$ &    5.22  & & $ -5.02\pm0.01$ &$1.84\pm0.02$ & 6.80   & \multicolumn{2}{c}{                    }\\
\multicolumn{2}{c}{V350 Ori  } &     $  4.39\pm0.06$ & $1.37\pm0.16$ &  0.69  & &                 &               &       & &   $   3.70\pm0.04 $ & $3.25\pm0.06$ &    1.84  & & $  3.68\pm0.05$ &$3.78\pm0.41$ & 1.01   & \multicolumn{2}{c}{ blue profile       }\\
\multicolumn{2}{c}{MWC 789   } &     $  2.57\pm0.08$ & $1.80\pm0.20$ &  0.58  & &                 &               &       & &   $   2.61\pm0.02 $ & $2.01\pm0.04$ &    1.95  & & $  2.62\pm0.05$ &$1.80\pm0.12$ & 1.12   & \multicolumn{2}{c}{ blue asy           }\\
\multicolumn{2}{c}{LkHA 208  } &     $ -0.13\pm0.04$ & $1.50\pm0.08$ &  1.30  & &                 &               &       & &   $   0.02\pm0.03 $ & $1.87\pm0.07$ &    3.01  & & $ -0.04\pm0.04$ &$1.21\pm0.13$ & 2.53   & \multicolumn{2}{c}{ two comp           }\\
\multicolumn{2}{c}{LkHA 339  } &     $ 11.57\pm0.02$ & $3.08\pm0.04$ &  2.97  & & $ 11.03\pm0.06$ & $2.27\pm0.14$ & 2.49  & &                     &               &          & &                 &              &        & \multicolumn{2}{c}{ self-abs           }\\
\multicolumn{2}{c}{LkHA 215  } &     $  2.67\pm0.02$ & $1.96\pm0.05$ &  1.70  & & $  2.34\pm0.03$ & $1.10\pm0.07$ & 2.10  & &   $   2.82\pm0.02 $ & $2.55\pm0.04$ &    6.62  & & $  2.89\pm0.05$ &$2.48\pm0.13$ & 7.62   & \multicolumn{2}{c}{ red-asy            }\\
\multicolumn{2}{c}{R Mon     } &     $  9.55\pm0.04$ & $1.57\pm0.09$ &  0.61  & &                 &               &       & &   $   9.51\pm0.02 $ & $2.35\pm0.06$ &    4.10  & & $  9.69\pm0.08$ &$1.93\pm0.07$ & 3.80   & \multicolumn{2}{c}{ red-asy            }\\
\multicolumn{2}{c}{V590 Mon  } &     $  5.48\pm0.04$ & $1.75\pm0.09$ &  0.72  & & $  4.96\pm0.12$ & $3.00\pm0.34$ & 0.84  & &   $   5.06\pm0.03 $ & $3.25\pm0.06$ &    2.41  & &                 &            &        & \multicolumn{2}{c}{ three comp         }\\
\multicolumn{2}{c}{          } &     $  9.02\pm0.05$ & $1.61\pm0.13$ &  0.68  & & $  8.76\pm0.10$ & $1.74\pm0.22$ & 0.83  & &   $   8.93\pm0.01 $ & $2.02\pm0.02$ &    6.37  & & $  8.95\pm0.04$ &$1.93\pm0.10$ & 6.74   & \multicolumn{2}{c}{                    }\\
\multicolumn{2}{c}{          } &     $ 11.48\pm0.06$ & $1.78\pm0.20$ &  0.56  & & $ 11.38\pm0.05$ & $1.31\pm0.15$ & 1.33  & &   $  11.45\pm0.01 $ & $1.58\pm0.02$ &    4.99  & & $ 11.53\pm0.04$ &$1.52\pm0.09$ & 5.87   & \multicolumn{2}{c}{                    }\\
\multicolumn{2}{c}{VV Ser    } &                     &               &        & &                 &               &       & &   $   5.34 $        & $1.73\pm0.04$ &    1.14  & & $  5.38\pm0.29$ &$1.43\pm0.29$ & 0.77   & \multicolumn{2}{c}{ three comp         }\\
\multicolumn{2}{c}{          } &                     &               &        & &                 &               &       & &   $   7.32\pm0.01 $ & $1.91\pm0.04$ &    1.29  & & $  7.53\pm0.29$ &$1.73\pm0.29$ & 1.92   & \multicolumn{2}{c}{                    }\\
\multicolumn{2}{c}{          } &                     &               &        & &                 &               &       & &   $   9.51\pm0.01 $ & $2.08\pm0.04$ &    0.92  & & $  9.13\pm0.29$ &$1.47\pm0.29$ & 0.81   & \multicolumn{2}{c}{                    }\\
\multicolumn{2}{c}{MWC 300   } &                     &               &        & &                 &               &       & &   $   6.54\pm0.21 $ & $2.12\pm0.21$ &    0.42  & & $  6.48\pm0.13$ &$0.55\pm0.21$ & 0.21   & \multicolumn{2}{c}{ three comp?        }\\
\multicolumn{2}{c}{          } &                     &               &        & &                 &               &       & &   $   8.27\pm0.21 $ & $1.87\pm0.21$ &    1.15  & & $  7.92\pm0.06$ &$1.44\pm0.20$ & 0.85   & \multicolumn{2}{c}{                    }\\
\multicolumn{2}{c}{          } &                     &               &        & &                 &               &       & &   $  10.12\pm0.21 $ & $1.62\pm0.21$ &    1.24  & & $ 10.00\pm0.08$ &$1.70\pm0.19$ & 0.76   & \multicolumn{2}{c}{                    }\\
\multicolumn{2}{c}{AS 310    } &                     &               &        & &                 &               &       & &   $   7.07\pm0.07 $ & $2.15\pm0.13$ &    0.53  & & $  7.28\pm0.20$ &$1.90\pm0.87$ & 0.54   & \multicolumn{2}{c}{                    }\\
\multicolumn{2}{c}{MWC 614   } &                     &               &        & &                 &               &       & &   $   2.22\pm0.09 $ & $1.10\pm0.20$ &    0.39  & &                 &              &        & \multicolumn{2}{c}{                    }\\
\multicolumn{2}{c}{Par 21    } &                     &               &        & &                 &               &       & &   $  16.79\pm0.04 $ & $1.07\pm0.09$ &    0.66  & & $ 16.97\pm0.07$ &$0.81\pm0.15$ & 0.82   & \multicolumn{2}{c}{                    }\\
\multicolumn{2}{c}{V1685 Cyg } &     $  7.61\pm0.02$ & $1.90\pm0.05$ &  5.22  & & $  7.79\pm0.03$ & $1.89\pm0.07$ & 6.60  & &   $   7.59\pm0.02 $ & $2.83\pm0.06$ &    9.98  & & $  7.60\pm0.07$ &$2.68\pm0.22$ & 9.83   & \multicolumn{2}{c}{ two comp           }\\
\multicolumn{2}{c}{          } &                     &               &        & &                 &               &       & &   $  12.64\pm0.08 $ & $3.27\pm0.19$ &    2.86  & &                 &              &        & \multicolumn{2}{c}{                    }\\
\multicolumn{2}{c}{Par 22    } &                     &               &        & &                 &               &       & &   $   4.84 $        & $3.17\pm0.02$ &    8.04  & & $  5.00\pm0.02$ &$2.90\pm0.04$ & 7.69   & \multicolumn{2}{c}{ two comp?          }\\
\multicolumn{2}{c}{          } &                     &               &        & &                 &               &       & &   $   9.21\pm0.03 $ & $2.90\pm0.06$ &    1.61  & & $  9.51\pm0.11$ &$3.14\pm0.33$ & 1.34   & \multicolumn{2}{c}{                    }\\
\multicolumn{2}{c}{PV Cep    } &     $  2.78\pm0.07$ & $0.66\pm0.16$ &  0.72  & &                 &               &       & &   $   2.15\pm0.03 $ & $2.98\pm0.06$ &    1.97  & &                 &              &        & \multicolumn{2}{c}{ red asy            }\\
\multicolumn{2}{c}{AS 442    } &     $  0.14\pm0.04$ & $1.12\pm0.09$ &  0.77  & &                 &               &       & &   $   0.23\pm0.01 $ & $1.44\pm0.04$ &    3.14  & & $  0.22\pm0.03$ &$1.43\pm0.09$ & 3.29   & \multicolumn{2}{c}{ red wing           }\\
\multicolumn{2}{c}{LkHA 134  } &                     &               &        & &                 &               &       & &   $   0.70\pm0.02 $ & $1.79\pm0.04$ &    2.15  & & $  0.94\pm0.03$ &$1.28\pm0.08$ & 1.89   & \multicolumn{2}{c}{                    }\\
\multicolumn{2}{c}{HD 200775 } &     $  2.30\pm0.02$ & $1.90\pm0.05$ &  1.49  & & $  2.66\pm0.14$ & $2.68\pm0.38$ & 0.95  & &   $   1.68\pm0.01 $ & $3.53\pm0.01$ &    4.71  & & $  1.60\pm0.03$ &$3.18\pm0.06$ & 4.91   & \multicolumn{2}{c}{ blue asy           }\\
\multicolumn{2}{c}{LkHA 324  } &                     &               &        & &                 &               &       & &   $  -2.43\pm0.03 $ & $5.62\pm0.07$ &    3.36  & & $ -2.17\pm0.04$ &$5.13\pm0.09$ & 3.96   & \multicolumn{2}{c}{ blue asy?          }\\
\multicolumn{2}{c}{V645 Cyg  } &     $-44.05\pm0.07$ & $2.94\pm0.15$ &  1.96  & & $-43.52\pm0.10$ & $1.45\pm0.19$ & 3.89  & &   $ -44.07\pm0.03 $ & $3.87\pm0.07$ &    3.64  & & $-43.88\pm0.03$ &$2.88\pm0.07$ & 4.93   & \multicolumn{2}{c}{                    }\\
\multicolumn{2}{c}{LkHA 234  } &     $-10.40\pm0.01$ & $2.15\pm0.02$ &  5.14  & & $-10.55\pm0.03$ & $1.98\pm0.08$ & 6.71  & &   $ -10.11\pm0.02 $ & $3.85\pm0.05$ &    10.58 & & $-10.11\pm0.02$ &$2.94\pm0.05$ & 10.21  & \multicolumn{2}{c}{ red wing?          }\\
\multicolumn{2}{c}{AS 477    } &                     &               &        & &                 &               &       & &   $   6.54\pm0.01 $ & $2.74\pm0.04$ &    3.83  & & $  6.66\pm0.03$ &$2.12\pm0.10$ & 4.41   & \multicolumn{2}{c}{ red asy?           }\\
\multicolumn{2}{c}{BD46      } &     $  6.25\pm0.02$ & $1.39\pm0.05$ &  2.21  & & $  6.89\pm0.07$ & $1.21\pm0.16$ & 2.37  & &   $   6.70\pm0.01 $ & $2.30\pm0.04$ &    3.54  & & $  6.71\pm0.03$ &$1.97\pm0.07$ & 3.44   & \multicolumn{2}{c}{ blue wing          }\\
\multicolumn{2}{c}{BH Cep    } &                     &               &        & &                 &               &       & &   $   1.81\pm0.07 $ & $0.93\pm0.16$ &    0.35  & &                 &              &        & \multicolumn{2}{c}{                    }\\
\multicolumn{2}{c}{V375 Lac  } &                     &               &        & &                 &               &       & &   $  -0.18 $        & $1.72\pm0.01$ &    7.13  & & $ -0.28$ &$1.55\pm0.01$ & 9.90   & \multicolumn{2}{c}{                    }\\
\multicolumn{2}{c}{IL Cep    } &     $ -9.99\pm0.03$ & $1.78\pm0.08$ &  1.62  & &                 &               &       & &   $ -10.13\pm0.04 $ & $2.95\pm0.09$ &    1.72  & & $ -9.95\pm0.09$ &$2.10\pm0.24$ & 0.83   & \multicolumn{2}{c}{ flat top           }\\
\multicolumn{2}{c}{MWC 1080  } &     $-30.52\pm0.02$ & $4.47\pm0.06$ &  1.91  & & $-30.52\pm0.05$ & $3.48\pm0.14$ & 1.46  & &                     &               &          & &                 &              &        & \multicolumn{2}{c}{ red profile,wings  }\\
\multicolumn{2}{c}{LkHA 259  } &     $ -7.13\pm0.06$ & $1.98\pm0.13$ &  1.17  & & $ -6.04\pm0.07$ & $1.45\pm0.17$ & 2.27  & &                     &               &          & &                 &              &        & \multicolumn{2}{c}{ blue profile       }\\
\enddata

\end{deluxetable}

\clearpage

\begin{deluxetable}{ccrrrrrrrrrrrrrrcrl}
\tabletypesize{\scriptsize} \tablecaption{Derived parameters of the
lines} \tablewidth{0pt} \tablehead{\multicolumn{2}{c}{Name} &
V$_{LSR}$ & $\frac{^{12}co(2-1)}{^{13}co(2-1)}$ & $\tau_{13co(2-1)}$
& $\tau_{12co(2-1)}$ & T$_{ex}$ & $\Theta_{s}$ &
$\frac{^{12}co(3-2)}{^{12}co(2-1)}$
& \multicolumn{2}{c}{N$_{H_{2}}$}\\
\multicolumn{2}{c}{} & \colhead{(km~s$^{-1}$)} & \colhead{} &
\colhead{} & \colhead{} &\colhead{(K)} & \colhead{$(\arcsec$)} &
\colhead{} &\multicolumn{2}{c}{($10^{21}$cm$^{-2}$)}   } \startdata
\multicolumn{2}{c}{MacC H12  } & -4.7  &3.6  &0.33 & 28.96  & 20.93   & 179  & 0.95   &    5.49   \\
\multicolumn{2}{c}{LkHA 198  } & -0.2  &3.1  &0.39 & 34.66  & 15.76   & 530  & 0.68   &    5.89   \\
\multicolumn{2}{c}{RNO 6     } & -36.0 &2.8  &0.44 & 39.32  & 10.20   & 230  & 0.77   &    3.06   \\
\multicolumn{2}{c}{XY Per    } & -4.2  &7.4  &0.15 & 12.92  & 11.24   & 133  & 0.64   &    1.55   \\
\multicolumn{2}{c}{V892 Tau  } & 7.2   &4.2  &0.27 & 24.20  & 12.05   & 310  & 0.64   &    2.76   \\
\multicolumn{2}{c}{AB Aur    } & 6.1   &     &     & 26.60  & 15.29   & 188  & 0.72   &    1.40   \\
\multicolumn{2}{c}{T Ori     } & 7.5   &4.4  &0.26 & 22.95  & 86.79   & 74   & 0.87   &    39.81  \\
\multicolumn{2}{c}{          } & 11.0  &6.4  &0.17 & 15.12  & 44.65   & 141  & 0.87   &    13.47  \\
\multicolumn{2}{c}{          } & 13.2  &8.9  &0.12 & 10.61  & 32.99   & 162  & 0.92   &    6.16   \\
\multicolumn{2}{c}{V380 Ori  } & 7.0   &     &     & 26.60  &         &      &        &           \\
\multicolumn{2}{c}{          } & 9.0   &2.2  &0.61 & 53.95  & 16.36   & 800  & 0.69   &    12.83  \\
\multicolumn{2}{c}{V586 Ori  } & 6.5   &     &     & 26.60  & 8.36    & 300  & 0.72   &    1.23   \\
\multicolumn{2}{c}{          } & 8.7   &     &     & 26.60  & 24.12   & 153  & 0.82   &    3.68   \\
\multicolumn{2}{c}{BF Ori    } & 6.2   &4.1  &0.28 & 24.88  & 17.12   & 285  & 0.72   &    4.42   \\
\multicolumn{2}{c}{          } & 9.2   &3.4  &0.35 & 31.00  & 14.57   & 164  & 1.00   &    3.59   \\
\multicolumn{2}{c}{          } & 10.9  &2.6  &0.49 & 43.21  & 7.55    & 300  & 0.54   &    1.66   \\
\multicolumn{2}{c}{Haro 13A  } & 5.6   &4.3  &0.26 & 23.56  & 14.25   & 600  & 0.68   &    4.11   \\
\multicolumn{2}{c}{V599 Ori  } & 5.0   &5.2  &0.21 & 19.01  & 11.43   &      & 0.70   &    2.66   \\
\multicolumn{2}{c}{          } & 7.2   &     &     & 26.60  &         &      &        &           \\
\multicolumn{2}{c}{RR Tau    } & -5.4  &4.5  &0.25 & 22.37  & 21.09   & 126  & 0.90   &    3.43   \\
\multicolumn{2}{c}{V350 Ori  } & 4.4   &6.3  &0.17 & 15.38  & 6.84    &      & 0.60   &    1.20   \\
\multicolumn{2}{c}{MWC 789   } & 2.6   &3.8  &0.31 & 27.18  & 7.06    &      & 0.50   &    1.33   \\
\multicolumn{2}{c}{LkHA 208  } & -0.1  &2.9  &0.42 & 37.63  & 10.05   & 290  & 0.45   &    2.28   \\
\multicolumn{2}{c}{LkHA 339  } & 11.3  &3.8  &0.31 & 27.18  & $>12.7$    &   &        &    $>7.53$   \\
\multicolumn{2}{c}{LkHA 215  } & 2.5   &5.1  &0.22 & 19.42  & 20.66   & 170  & 0.86   &    4.08   \\
\multicolumn{2}{c}{R Mon     } & 9.6   &10.1 &0.10 & 9.28   & 12.58   & 250  & 0.61   &    0.91   \\
\multicolumn{2}{c}{V590 Mon  } & 5.2   &6.2  &0.18 & 15.65  & 10.16   &      &        &    1.08   \\
\multicolumn{2}{c}{          } & 8.9   &11.8 &0.09 & 7.88   & 18.24   & 207  & 0.83   &    1.09   \\
\multicolumn{2}{c}{          } & 11.4  &7.9  &0.14 & 12.05  & 17.88   & 152  & 0.82   &    1.26   \\
\multicolumn{2}{c}{VV Ser    } & 5.4   &     &     & 26.60  & 6.23    & 270  & 0.44   &    0.98   \\
\multicolumn{2}{c}{          } & 7.4   &     &     & 26.60  & 12.69   & 74   & 0.71   &    2.04   \\
\multicolumn{2}{c}{          } & 9.3   &     &     & 26.60  & 6.81    & 140  & 0.43   &    1.25   \\
\multicolumn{2}{c}{MWC 300   } & 6.5   &     &     & 26.60  & 4.32    & 300  & 0.09   &    1.00   \\
\multicolumn{2}{c}{          } & 8.1   &     &     & 26.60  & 6.53    & 220  & 0.51   &    1.08   \\
\multicolumn{2}{c}{          } & 10.1  &     &     & 26.60  & 6.08    & 500  & 0.58   &    0.91   \\
\multicolumn{2}{c}{AS 310    } & 7.2   &     &     & 26.60  & 6.47    & 95   & 0.55   &    1.24   \\
\multicolumn{2}{c}{MWC 614   } & 2.2   &     &     & 26.60  & 4.56    &      &        &    0.35   \\
\multicolumn{2}{c}{Par 21    } & 16.9  &     &     & 26.60  & 8.18    & 78   & 0.50   &    0.74   \\
\multicolumn{2}{c}{V1685 Cyg } & 7.7   &2.8  &0.44 & 39.32  & 22.23   & 330  & 0.83   &    10.00  \\
\multicolumn{2}{c}{          } & 12.6  &     &     & 26.60  & 11.26   &      &        &    2.00   \\
\multicolumn{2}{c}{Par 22    } & 4.9   &     &     & 26.60  & 18.75   & 340  & 0.83   &    5.64   \\
\multicolumn{2}{c}{          } & 9.4   &     &     & 26.60  & 7.84    & 200  & 0.74   &    1.92   \\
\multicolumn{2}{c}{PV Cep    } & 2.8   &12.4 &0.08 & 7.48   & 9.05    &      &        &    0.43   \\
\multicolumn{2}{c}{AS 442    } & 0.1   &5.2  &0.21 & 19.01  & 12.40   & 165  & 0.77   &    1.11   \\
\multicolumn{2}{c}{LkHA 134  } & 0.8   &     &     & 26.60  & 9.07    & 205  & 0.50   &    1.34   \\
\multicolumn{2}{c}{HD 200775 } & 2.5   &5.9  &0.19 & 16.53  & 15.13   & 193  & 0.73   &    3.08   \\
\multicolumn{2}{c}{LkHA 324  } & -2.3  &     &     & 26.60  & 14.55   & 135  & 0.71   &    7.11   \\
\multicolumn{2}{c}{V645 Cyg  } & -43.8 &2.4  &0.54 & 47.97  & 18.13   & 109  & 0.65   &    12.32  \\
\multicolumn{2}{c}{LkHA 234  } & -10.5 &3.7  &0.32 & 28.04  & 22.56   & 390  & 0.66   &    10.13  \\
\multicolumn{2}{c}{AS477/BD46} & 6.6   &2.7  &0.46 & 41.17  & 15.10   & 147  & 0.59   &    4.68   \\
\multicolumn{2}{c}{BH Cep    } & 1.8   &     &     & 26.60  & 4.41    &      &        &    0.29   \\
\multicolumn{2}{c}{V375 Lac  } & -0.2  &     &     & 26.60  & 28.38   & 119  & 0.81   &    5.76   \\
\multicolumn{2}{c}{IL Cep    } & -10.0 &1.8  &0.81 & 72.17  & 6.51    &      & 0.30   &    5.34   \\
\multicolumn{2}{c}{MWC 1080  } & -30.5 &     &     & 26.60  & $>11.8$    &   &        &    $>7.19$   \\
\multicolumn{2}{c}{LkHA 259  } & -6.6  &     &     & 26.60  & $>6.4$     &   &        &    $>3.49$   \\
\enddata

\end{deluxetable}

\clearpage
\begin{deluxetable}{ccrrrrrrrrrrrrcrl}
\rotate \tabletypesize{\scriptsize} \setlength{\tabcolsep}{0.05in}
\tablecaption{The SED fitting results}
 \tablewidth{0pt} \tablehead{
  Name& A$_{v}$  &log(Age) &M$_{*}$ &R$_{*}$ &log(L$_{*}$) &log(T$_{*}$)  &log(M$_{env}$) & log$\dot{M}_{env}$) & log(M$_{disk})$ & Incl & log(R$_{out}$) & log($\dot{M}_{disk}$)\\

&(mag) &log(yr) & (M$_{\sun}$)  &(R$_{\sun}$)  & log(L$_{\sun}$) &
log(K) & log(M$_{\sun}$)  &log(M$_{\sun}$yr$^{-1}$) &
log(M$_{\sun}$) &($\arcdeg$) & log(AU) &log(M$_{\sun}$yr$^{-1}$) }
\startdata
MacC H12     &  $  0.61\pm0.57 $    &   $3.73\pm0.32$   &    $  1.84\pm0.16$   & $ 14.96\pm2.44  $    &   $ 1.79\pm0.13$    &  $3.62       $    & $   0.32\pm0.16$   & $-4.88\pm0.07$    & $-1.74\pm0.10$    & $  18.19         $   &  $1.21\pm0.32$     & $  -6.55\pm0.35 $  \\
LkHA 198     &  $  0.00        $    &   $3.07\pm0.06$   &    $  3.84\pm0.38$   & $ 29.95\pm5.60  $    &   $ 2.42\pm0.14$    &  $3.62       $    & $   0.06\pm0.35$   & $-4.56\pm0.13$    & $-2.12\pm0.63$    & $  57.17\pm24.78 $   &  $0.54\pm0.16$     & $  -5.33\pm0.12 $  \\
Vx Cas       &  $  1.48\pm0.30 $    &   $6.66\pm0.25$   &    $  3.62\pm0.24$   & $  2.22\pm0.08  $    &   $ 2.17\pm0.11$    &  $4.13\pm0.02$    & $  -4.30\pm0.54$   &                   & $-2.55\pm0.13$    & $  41.80\pm19.87 $   &  $3.00\pm0.32$     & $  -7.89\pm0.46 $  \\
RNO 6        &  $  0.71\pm0.65 $    &   $6.03\pm0.07$   &    $  5.10\pm0.42$   & $  2.76\pm0.08  $    &   $ 2.75\pm0.12$    &  $4.23\pm0.02$    & $   1.08\pm0.10$   & $-8.47\pm0.31$    & $-1.78\pm0.23$    & $  78.59\pm2.92  $   &  $2.40\pm0.21$     & $  -6.79\pm0.94 $  \\
IP Per       &  $  0.00\pm0.01 $    &   $5.71\pm0.09$   &    $  2.17\pm0.47$   & $  5.03\pm0.50  $    &   $ 1.09\pm0.09$    &  $3.67\pm0.01$    & $  -1.64\pm0.44$   & $-5.62\pm0.41$    & $-1.49\pm0.35$    & $  28.66\pm9.47  $   &  $2.45\pm0.25$     & $  -6.85\pm0.78 $  \\
V892 Tau     &  $  6.44\pm1.95 $    &   $6.50\pm0.37$   &    $  2.50\pm0.88$   & $  4.04\pm2.84  $    &   $ 1.81\pm0.24$    &  $3.98\pm0.18$    & $  -0.88\pm0.38$   & $-5.62\pm0.38$    & $-1.29\pm0.30$    & $  30.50\pm9.53  $   &  $2.69\pm0.35$     & $  -7.14\pm0.15 $  \\
XY Per       &  $  3.06\pm0.09 $    &   $6.99       $   &    $  2.81       $   & $  1.93         $    &   $ 1.75       $    &  $4.06       $    & $  -5.71       $   &                   & $-2.02$           & $  78.61\pm2.92  $   &  $2.77$     & $  -8.81 $  \\
AB Aur       &  $  1.04        $    &   $5.14       $   &    $  1.10       $   & $  6.73         $    &   $ 1.10       $    &  $3.62       $    & $  -1.33       $   & $-5.65$           & $-2.12$           & $  31.79         $   &  $2.15$     & $  -7.34 $  \\
MWC 480       &  $  0.33\pm0.35 $    &   $6.33\pm0.16$   &    $  3.04\pm0.33$   & $  4.47\pm1.10  $    &   $ 1.78\pm0.92$    &  $3.86\pm0.19$    & $  -6.13\pm0.35$   &                   & $-1.29\pm0.25$    & $  54.75\pm19.35 $   &  $2.38\pm0.25$     & $  -6.36\pm0.67 $  \\
HD 35929     &  $  0.24\pm0.16 $    &   $6.36\pm0.19$   &    $  3.10\pm0.47$   & $  5.19\pm0.95  $    &   $ 1.81\pm0.23$    &  $3.85\pm0.03$    & $  -2.87\pm0.36$   &                   & $-4.87\pm0.63$    & $  56.30\pm19.64 $   &  $3.61\pm0.72$     & $ -10.66\pm0.71 $  \\
HD 36112     &  $  0.60\pm0.04 $    &   $6.96\pm0.05$   &    $  1.95\pm0.05$   & $  1.83\pm0.05  $    &   $ 1.11\pm0.02$    &  $3.91       $    & $  -5.86\pm0.44$   &                   & $-1.78\pm0.49$    & $  35.05\pm11.45 $   &  $2.58\pm0.33$     & $  -8.23\pm0.14 $  \\
HD 245185    &  $  0.00        $    &   $6.13       $   &    $  3.74       $   & $  5.68         $    &   $ 2.17       $    &  $3.93       $    & $  -6.84       $   &                   & $-1.41$           & $  81.37         $   &  $2.29$     & $  -7.19 $  \\
T Ori        &  $  1.47\pm0.16 $    &   $6.69\pm0.28$   &    $  3.72\pm0.51$   & $  2.33\pm0.31  $    &   $ 2.27\pm0.18$    &  $4.13\pm0.04$    & $  -5.42\pm0.46$   &                   & $-1.26\pm0.51$    & $  53.80\pm15.54 $   &  $2.60\pm0.71$     & $  -6.54\pm0.43 $  \\
CQ Tau       &  $  2.31\pm0.13 $    &   $6.85\pm0.15$   &    $  2.82\pm0.27$   & $  2.07\pm0.03  $    &   $ 1.78\pm0.16$    &  $4.04\pm0.04$    & $  -4.50\pm0.32$   &                   & $-2.03\pm0.47$    & $  47.41\pm31.50 $   &  $2.67\pm0.23$     & $  -7.47\pm0.54 $  \\
V380 Ori     &  $  2.87\pm1.42 $    &   $5.87\pm0.11$   &    $  4.68\pm0.05$   & $  6.62\pm2.95  $    &   $ 2.62\pm0.21$    &  $4.03\pm0.13$    & $   0.88\pm0.21$   & $-5.66\pm0.43$    & $-3.51\pm0.29$    & $  45.12\pm18.54 $   &  $2.88\pm0.39$     & $  -8.52\pm0.48 $  \\
V586 Ori     &  $  1.00        $    &   $6.01       $   &    $  3.86       $   & $  7.62         $    &   $ 1.93       $    &  $3.80       $    & $  -0.06       $   & $-7.56$           & $-1.60$           & $  81.37         $   &  $2.24$     & $  -6.86 $  \\
BF Ori       &  $  1.81\pm0.31 $    &   $6.72\pm0.17$   &    $  3.09\pm0.31$   & $  2.19\pm0.29  $    &   $ 1.94\pm0.16$    &  $4.07\pm0.04$    & $  -4.60\pm0.98$   &                   & $-2.94\pm0.57$    & $  49.03\pm21.24 $   &  $2.58\pm0.53$     & $  -8.08\pm0.59 $  \\
HD37411      &  $ 11.65        $    &   $6.63       $   &    $  0.35       $   & $  1.04         $    &   $-0.15       $    &  $3.55       $    & $  -8.77       $   &                   & $-2.33$           & $  63.26         $   &  $2.09$     & $  -7.18 $  \\
Haro 13A     &  $  0.00        $    &   $3.02       $   &    $  3.47       $   & $ 24.48         $    &   $ 1.34\pm0.32$    &  $3.63       $    & $  -0.25\pm0.65$   & $-4.71$           & $-1.84$           & $  64.16\pm21.08 $   &  $0.67$     & $  -5.62\pm0.36 $  \\
V599 Ori     &  $  2.94\pm1.62 $    &   $5.64\pm0.30$   &    $  1.83\pm1.19$   & $  5.29\pm1.59  $    &   $ 1.20\pm0.31$    &  $3.65\pm0.06$    & $  -1.36\pm0.52$   & $-5.76\pm0.35$    & $-1.24\pm0.27$    & $  42.47\pm24.22 $   &  $2.59\pm0.46$     & $  -6.71\pm0.37 $  \\
RR Tau       &  $  1.82\pm0.27 $    &   $6.53\pm0.28$   &    $  3.68\pm0.42$   & $  2.60\pm0.49  $    &   $ 2.30\pm0.17$    &  $4.12\pm0.03$    & $  -5.71\pm0.46$   &                   & $-1.12\pm0.24$    & $  43.72\pm19.52 $   &  $2.60\pm0.32$     & $  -6.41\pm0.32 $  \\
V350 Ori     &  $  1.69\pm0.47 $    &   $6.73\pm0.16$   &    $  2.73\pm0.39$   & $  2.05\pm0.31  $    &   $ 1.75\pm0.19$    &  $4.03\pm0.06$    & $  -3.48\pm0.39$   &                   & $-2.52\pm0.50$    & $  43.21\pm19.30 $   &  $3.39\pm0.32$     & $  -7.29\pm0.73 $  \\
MWC 789      &  $  2.42\pm0.77 $    &   $6.24\pm0.17$   &    $  3.90\pm0.37$   & $  2.81\pm0.29  $    &   $ 2.36\pm0.10$    &  $4.12\pm0.02$    & $  -6.47\pm0.18$   &                   & $-1.02\pm0.13$    & $  22.65\pm6.39  $   &  $2.43\pm0.05$     & $  -6.49\pm0.37 $  \\
LkHA 208     &  $  5.39\pm0.09 $    &   $5.97       $   &    $  4.23       $   & $  6.69         $    &   $ 2.32       $    &  $3.93       $    & $  -0.93       $   & $-7.33$           & $-1.00$           & $  55.80\pm14.37 $   &  $3.06$     & $  -7.71 $  \\
LkHA 339     &  $  2.96\pm0.12 $    &   $6.60\pm0.31$   &    $  3.42\pm0.60$   & $  2.25\pm0.25  $    &   $ 2.17\pm0.28$   &   $ 4.11\pm0.05$   & $  -2.38\pm0.76$    &                  & $-1.78\pm0.36$    & $  40.20\pm21.08 $   &  $3.73\pm0.52$     & $  -8.67\pm0.38 $  \\
LkHA 215     &  $  1.34\pm0.35 $    &   $5.74       $   &    $  5.13       $   & $  8.46         $    &   $ 2.59       $    &  $3.95       $    & $   1.05       $   & $-5.01$           & $-2.15$           & $  46.40\pm21.09 $   &  $3.65$     & $  -9.19 $  \\
R Mon        &  $  4.87        $    &   $4.24       $   &    $  4.67       $   & $ 24.59         $    &   $ 2.29       $    &  $3.64       $    & $  -0.61       $   & $-4.56$           & $-1.95$           & $  18.19         $   &  $1.55$     & $  -6.97 $  \\
V590 Mon     &  $  0.42\pm0.15 $    &   $6.45\pm0.54$   &    $  4.71\pm0.20$   & $  5.86\pm3.12  $    &   $ 2.47\pm0.04$    &  $4.06\pm0.15$    & $  -1.86\pm0.68$   &  $-6.57\pm0.73$   & $-2.25\pm0.31$    & $  84.45\pm2.87  $   &  $3.03\pm0.36$     & $  -8.03\pm0.31 $  \\
GU CMa       &  $  0.42        $    &   $6.90       $   &    $  3.14       $   & $  2.05         $    &   $ 1.93       $    &  $4.09       $    & $  -3.51       $   &                   & $-7.11$           & $  49.46         $   &  $3.62$     & $ -13.76 $  \\
HD 141569    &  $  0.26\pm0.04 $    &   $6.93       $   &    $  2.14       $   & $  1.74         $    &   $ 1.33       $    &  $3.97       $    & $  -2.99       $   &                   & $-4.81$           & $  56.31\pm23.40 $   &  $3.72$     & $ -11.82 $  \\
VV Ser       &  $  3.98\pm0.25 $    &   $6.81\pm0.20$   &    $  3.32\pm0.35$   & $  2.16\pm0.15  $    &   $ 2.10\pm0.18$    &  $4.10\pm0.03$    & $  -5.83\pm0.63$   &                   & $-1.39\pm0.31$    & $  45.71\pm15.48 $   &  $2.72\pm0.28$     & $  -6.43\pm0.34 $  \\
MWC 300      &  $  4.23\pm2.13 $    &   $6.56\pm0.11$   &    $  7.37\pm0.67$   & $  3.33\pm0.19  $    &   $ 3.35\pm0.13$    &  $4.33\pm0.03$    & $  -6.44\pm0.38$   &                   & $-1.05\pm0.01$    & $  62.27\pm13.03 $   &  $2.27\pm0.23$     & $  -5.80\pm0.80 $  \\
AS 310       &  $  6.65\pm3.01 $    &   $5.85\pm0.21$   &    $  9.86\pm0.53$   & $  3.90\pm0.10  $    &   $ 3.79\pm0.09$    &  $4.41\pm0.01$    & $   1.61\pm0.08$   & $-4.31\pm0.9$     & $-0.59\pm0.11$    & $  87.13         $   &  $3.26\pm0.44$     & $  -5.53\pm0.80 $  \\
Par 21       &  $  1.61        $    &   $6.13       $   &    $  3.74       $   & $  5.68         $    &   $ 2.17       $    &  $3.93       $    & $  -6.84       $   &                   & $-1.41$           & $  87.13         $   &  $2.29$     & $  -7.19 $  \\
V1295 Aql    &  $  0.36        $    &   $6.69\pm0.30$   &    $  3.22\pm0.37$   & $  2.29\pm0.34  $    &   $ 2.00\pm0.18$    &  $4.07\pm0.01$    & $  -6.67\pm0.30$   &                   & $-2.55\pm0.31$    & $  66.45\pm9.44  $   &  $1.55\pm0.09$     & $  -7.18\pm0.53 $  \\
V1685 Cyg    &  $  1.35\pm0.95 $    &   $4.93\pm0.37$   &    $  6.75\pm1.40$   & $ 24.50\pm12.49 $    &   $ 3.08\pm0.14$    &  $3.85\pm0.13$    & $   1.30\pm0.24$   & $-4.08\pm0.32$    & $-1.04\pm0.35$    & $  35.64\pm14.64 $   &  $1.74\pm0.34$     & $  -4.61\pm0.38 $  \\
Par 22       &  $  3.67\pm2.57 $    &   $5.35\pm0.35$   &    $  1.29\pm1.14$   & $  6.46\pm2.88  $    &   $ 1.24\pm0.34$    &  $3.59\pm0.09$    & $  -0.06\pm0.48$   & $-4.21\pm0.55$    & $-1.71\pm0.35$    & $  54.82\pm29.65 $   &  $2.00\pm1.47$     & $  -6.35\pm0.52 $  \\
PV Cep       &  $  3.16\pm0.43 $    &   $3.08\pm0.06$   &    $  0.28\pm0.09$   & $  5.69\pm0.25  $    &   $ 1.77\pm0.18$    &  $3.51\pm0.03$    & $  -0.49\pm0.19$   & $-5.33\pm0.06$    & $-1.53\pm0.34$    & $  18.19         $   &  $0.30\pm0.16$     & $  -4.40\pm0.06 $  \\
AS 442       &  $  1.98\pm0.24 $    &   $6.54\pm0.22$   &    $  4.31\pm0.37$   & $  2.67\pm0.60  $    &   $ 2.48\pm0.15$    &  $4.17\pm0.04$    & $  -3.41\pm0.61$   &                   & $-1.99\pm0.39$    & $  55.67\pm13.68 $   &  $2.92\pm0.45$     & $  -6.99\pm0.47 $  \\
LkHA 134     &  $  2.67\pm0.11 $    &   $6.27\pm0.13$   &    $  4.13\pm0.31$   & $  2.67\pm0.36  $    &   $ 2.45\pm0.14$    &  $4.16\pm0.02$    & $  -2.07\pm0.06$   & $-8.05\pm0.50$    & $-1.75\pm0.26$    & $  58.33\pm20.25 $   &  $3.90\pm0.20$     & $  -9.06\pm0.52 $  \\
HD 200775    &  $  0.12\pm0.08 $    &   $5.34\pm0.01$   &    $  6.65\pm0.09$   & $ 14.22\pm0.06  $    &   $ 2.78\pm0.03$    &  $3.88\pm0.01$    & $   1.61\pm0.05$   & $-3.59\pm0.30$    & $-2.33\pm0.01$    & $  26.17\pm11.03 $   &  $1.93\pm0.12$     & $  -7.86\pm0.11 $  \\
LkHA 324     &  $  1.01\pm1.09 $    &   $6.16\pm1.27$   &    $  2.32\pm0.77$   & $  4.87\pm1.24  $    &   $ 1.47\pm0.43$    &  $3.76\pm0.20$    & $  -1.50\pm0.51$   & $-5.54\pm0.42$    & $-1.89\pm0.45$    & $  41.68\pm19.41 $   &  $2.67\pm0.97$     & $  -7.74\pm0.56 $  \\
HD 203024    &  $  0.45\pm0.30 $    &   $6.89\pm0.07$   &    $  2.35\pm0.15$   & $  1.89\pm0.06  $    &   $ 1.48\pm0.09$    &  $3.99\pm0.03$    & $  -5.33\pm0.36$   &                   & $-2.06\pm0.02$    & $  74.29\pm4.39  $   &  $2.27\pm0.26$     & $  -8.20\pm0.30 $  \\
V645 Cyg     &  $  0.42        $    &   $5.16       $   &    $ 10.59       $   & $  4.16         $    &   $ 3.88       $    &  $4.42       $    & $   2.46       $   & $-3.38$           & $-2.26$           & $  18.19         $   &  $2.96$     & $  -7.44 $  \\
LkHA 234     &  $  3.05\pm3.29 $    &   $5.15\pm0.02$   &    $  8.96\pm0.25$   & $  4.91\pm0.39  $    &   $ 3.76\pm0.01$    &  $4.35\pm0.02$    & $   2.82\pm0.06$   & $-2.97\pm0.05$    & $-0.79\pm0.72$    & $  45.13\pm4.01  $   &  $2.21\pm0.47$     & $  -5.34\pm1.02 $  \\
AS 477       &  $  0.60\pm0.25 $    &   $6.43\pm0.22$   &    $  5.11\pm0.66$   & $  2.72\pm0.23  $    &   $ 2.76\pm0.20$    &  $4.22\pm0.04$    & $  -5.86\pm0.45$   &                   & $-1.65\pm0.34$    & $  59.44\pm13.66 $   &  $2.55\pm0.36$     & $  -6.27\pm0.35 $  \\
LkHA 257     &  $  2.62\pm0.83 $    &   $6.62\pm0.16$   &    $  3.61\pm0.64$   & $  2.68\pm1.60  $    &   $ 2.18\pm0.16$    &  $4.11\pm0.09$    & $  -2.31\pm0.66$   & $-6.99\pm0.70$    & $-2.46\pm0.71$    & $  57.05\pm22.40 $   &  $3.43\pm0.34$     & $  -7.14\pm0.71 $  \\
BH Cep       &  $  0.80\pm0.34 $    &   $6.61\pm0.29$   &    $  3.27\pm0.21$   & $  3.35\pm1.33  $    &   $ 2.05\pm0.13$    &  $4.03\pm0.06$    & $  -3.49\pm0.33$   &                   & $-2.50\pm0.70$    & $  84.26\pm2.88  $   &  $3.53\pm0.14$     & $  -8.24\pm0.96 $  \\
SV Cep       &  $  2.37\pm0.17 $    &   $6.71\pm0.17$   &    $  3.33\pm0.48$   & $  2.12\pm0.18  $    &   $ 2.07\pm0.24$    &  $4.10\pm0.04$    & $  -6.40\pm0.54$   &                   & $-2.32\pm0.42$    & $  48.85\pm22.50 $   &  $2.31\pm0.36$     & $  -7.55\pm0.68 $  \\
V375 Lac     &  $  1.85\pm1.14 $    &   $3.71\pm0.40$   &    $  1.88\pm0.51$   & $ 15.15\pm2.93  $    &   $ 1.92\pm0.16$    &  $3.62\pm0.01$    & $   0.55\pm0.79$   & $-4.45\pm0.15$    & $-1.28\pm0.75$    & $  18.19         $   &  $0.85\pm0.32$     & $  -5.18\pm0.48 $  \\
IL Cep       &  $  0.28\pm0.32 $    &   $6.02\pm0.10$   &    $  3.85\pm0.19$   & $  7.53\pm1.07  $    &   $ 1.94\pm0.15$    &  $3.81\pm0.07$    & $  -2.43\pm0.61$   & $-7.08\pm0.54$    & $-2.54\pm0.68$    & $  47.87\pm18.62 $   &  $2.47\pm0.45$     & $  -8.25\pm0.34 $  \\
MWC 1080     &  $  1.01\pm0.78 $    &   $5.37\pm0.02$   &    $ 10.59\pm0.87$   & $  4.00\pm0.19  $    &   $ 3.88\pm0.12$    &  $4.43\pm0.02$    & $   2.08\pm0.17$   & $-3.58\pm0.21$    & $-1.01\pm0.47$    & $  38.03\pm4.59  $   &  $2.72\pm0.30$     & $  -5.92\pm0.37 $  \\
LkHA 259     &  $  1.78\pm0.19 $    &   $3.29\pm0.26$   &    $  2.63\pm0.75$   & $ 22.94\pm5.25  $    &   $ 2.19\pm0.22$    &  $3.61       $    & $   1.02\pm0.15$   & $-3.09\pm0.14$    & $-1.69\pm0.32$    & $  18.19         $   &  $0.59\pm0.14$     & $  -5.50\pm0.28 $  \\

\enddata

\end{deluxetable}

\begin{figure}
\includegraphics[angle=0,scale=.50]{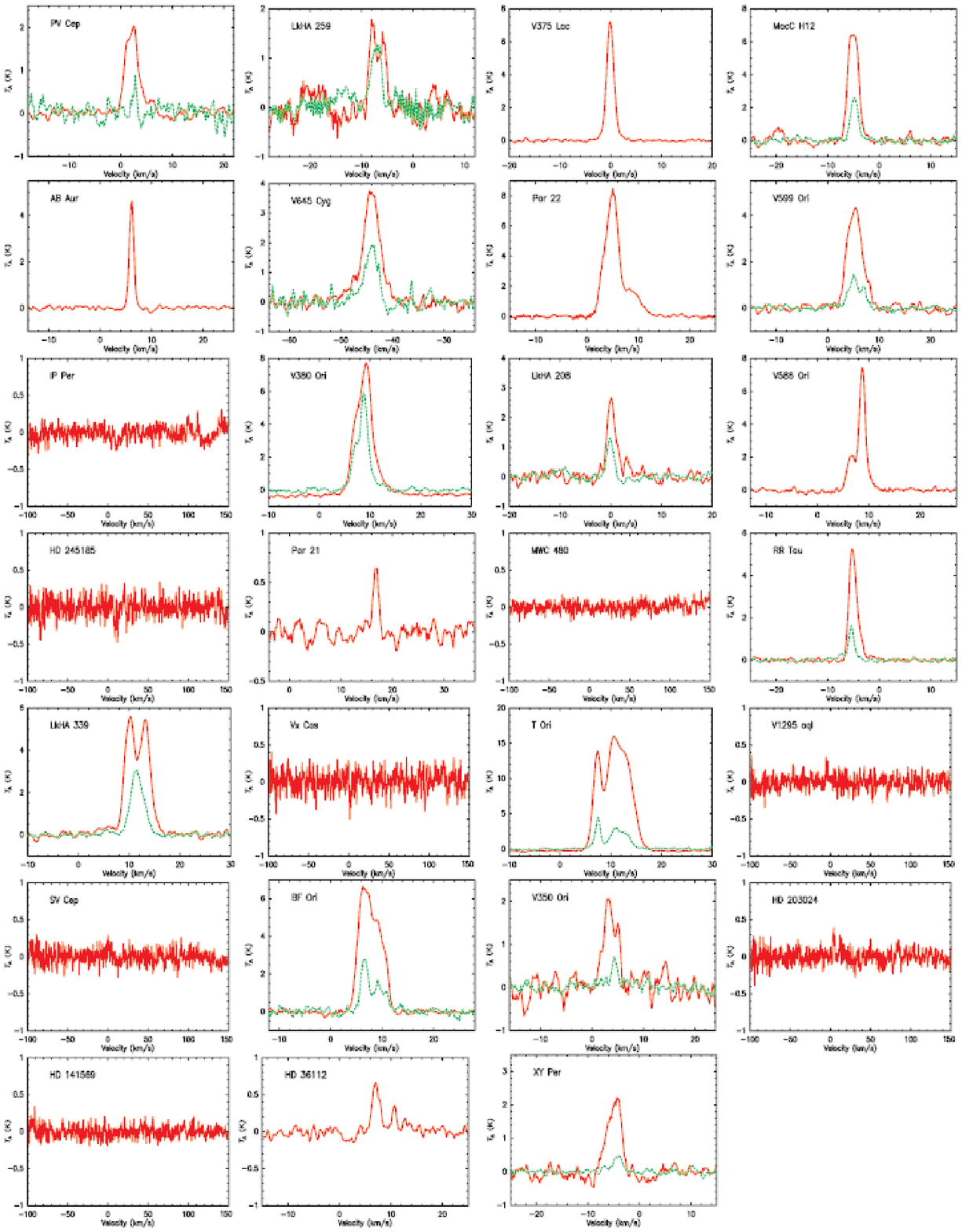}
\caption{$^{12}$CO (2-1) (red) and $^{13}$CO (2-1) (green) lines of
the A Type stars. The source names are plotted in the upper-left
corner of each panel.}
\end{figure}

\clearpage

\begin{figure}
\includegraphics[angle=0,scale=.50]{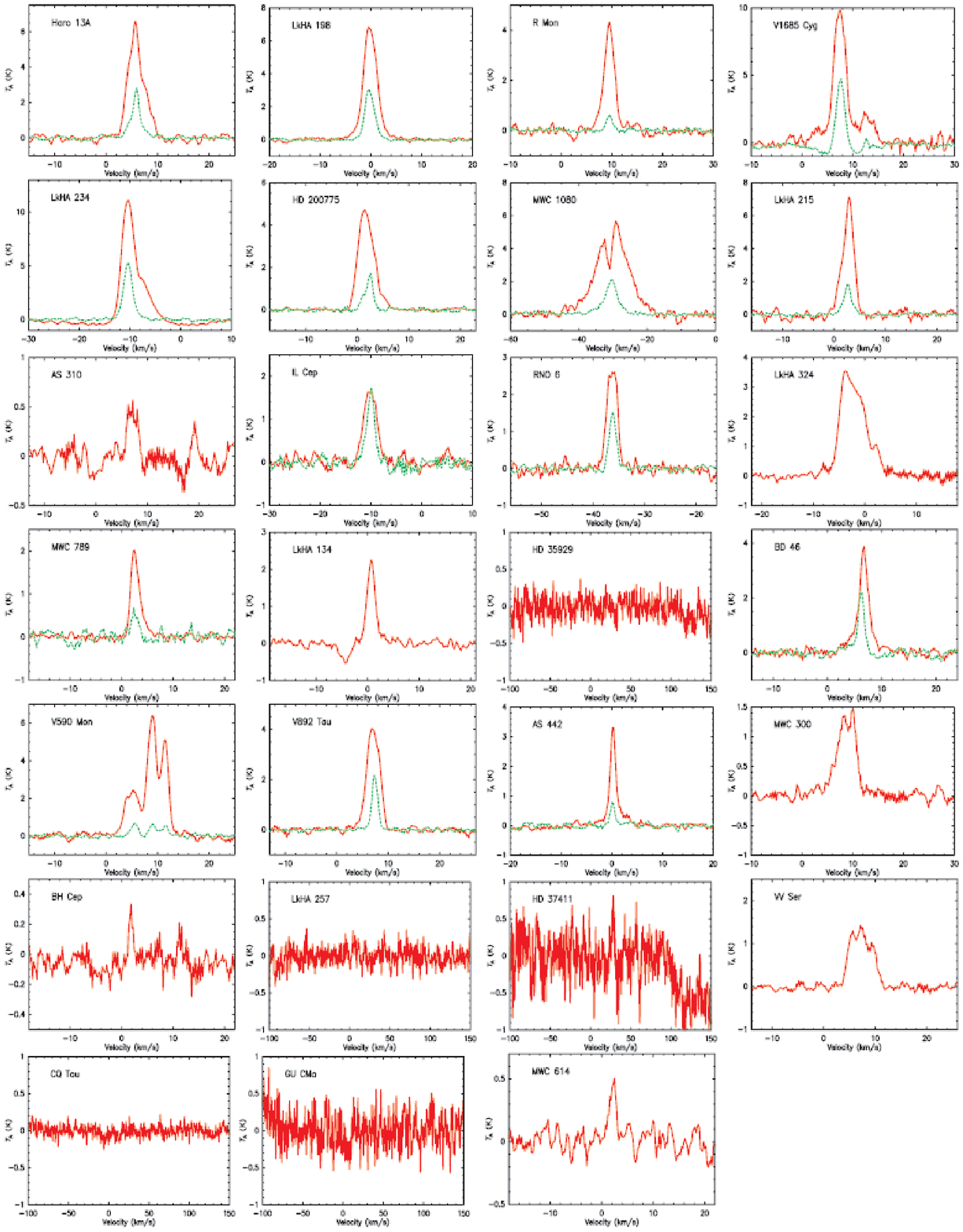}
\caption{$^{12}$CO (2-1) (red) and $^{13}$CO (2-1) (green) lines of
the B/F Type stars. The source names are plotted in the upper-left
corner of each panel.}
\end{figure}

\clearpage
\begin{figure}
\includegraphics[angle=0,scale=.50]{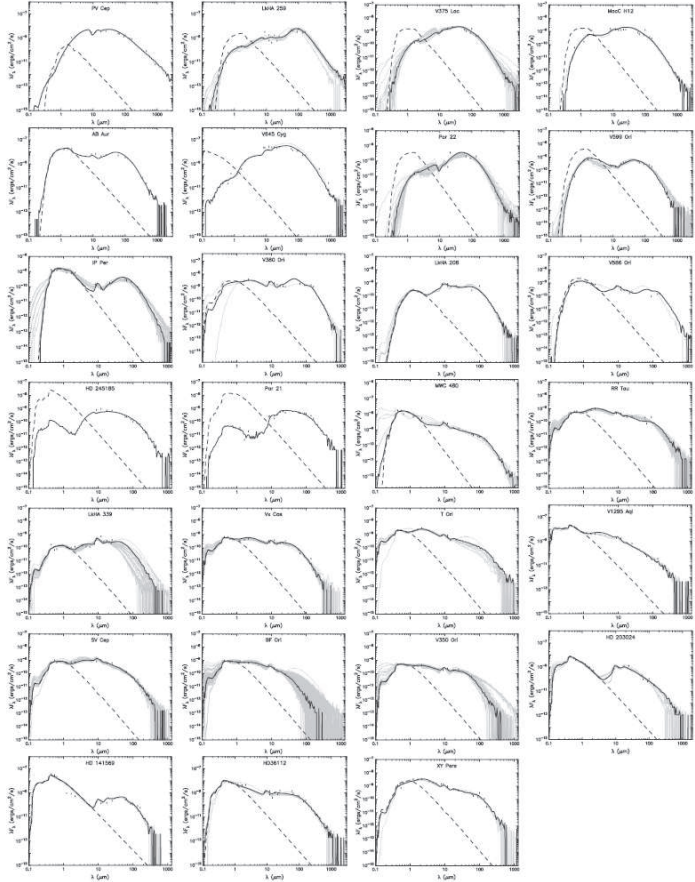}
\caption{The SEDs of A Type stars. The observed data are plotted as
circular symbols. The black solid line represents the best fit to
the data, while the grey solid lines represent the fits with
$\chi^{2}-\chi_{best}^{2}<3\times n_{data}$. The dashed lines
represent photospheric contributions, including the effect of
foreground extinction, from the best-fitting models. As shown by the
vertical lines, some model SEDs are with bad signal-to-noise ratios
for wavelengths beyond 1-100 $\micron$ \citep{ro06}.}
\end{figure}

\clearpage

\begin{figure}
\includegraphics[angle=0,scale=.50]{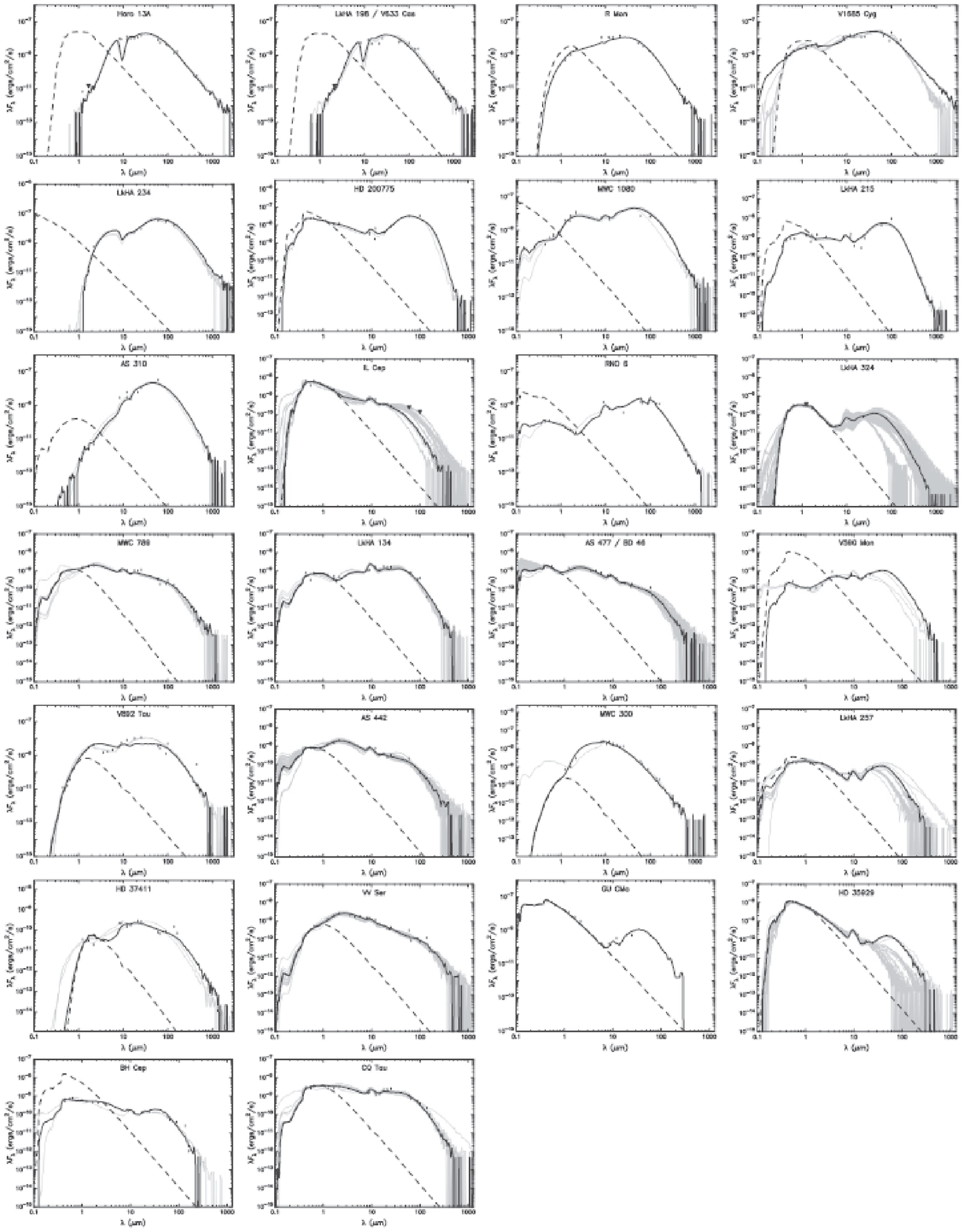}
\caption{The SEDs of  B/F Type stars. The symbols and lines are the
same as in Fig.3}
\end{figure}

\clearpage

\begin{figure}
\begin{minipage}[c]{0.5\textwidth}
  \centering
  \includegraphics[width=80mm,height=70mm,angle=0]{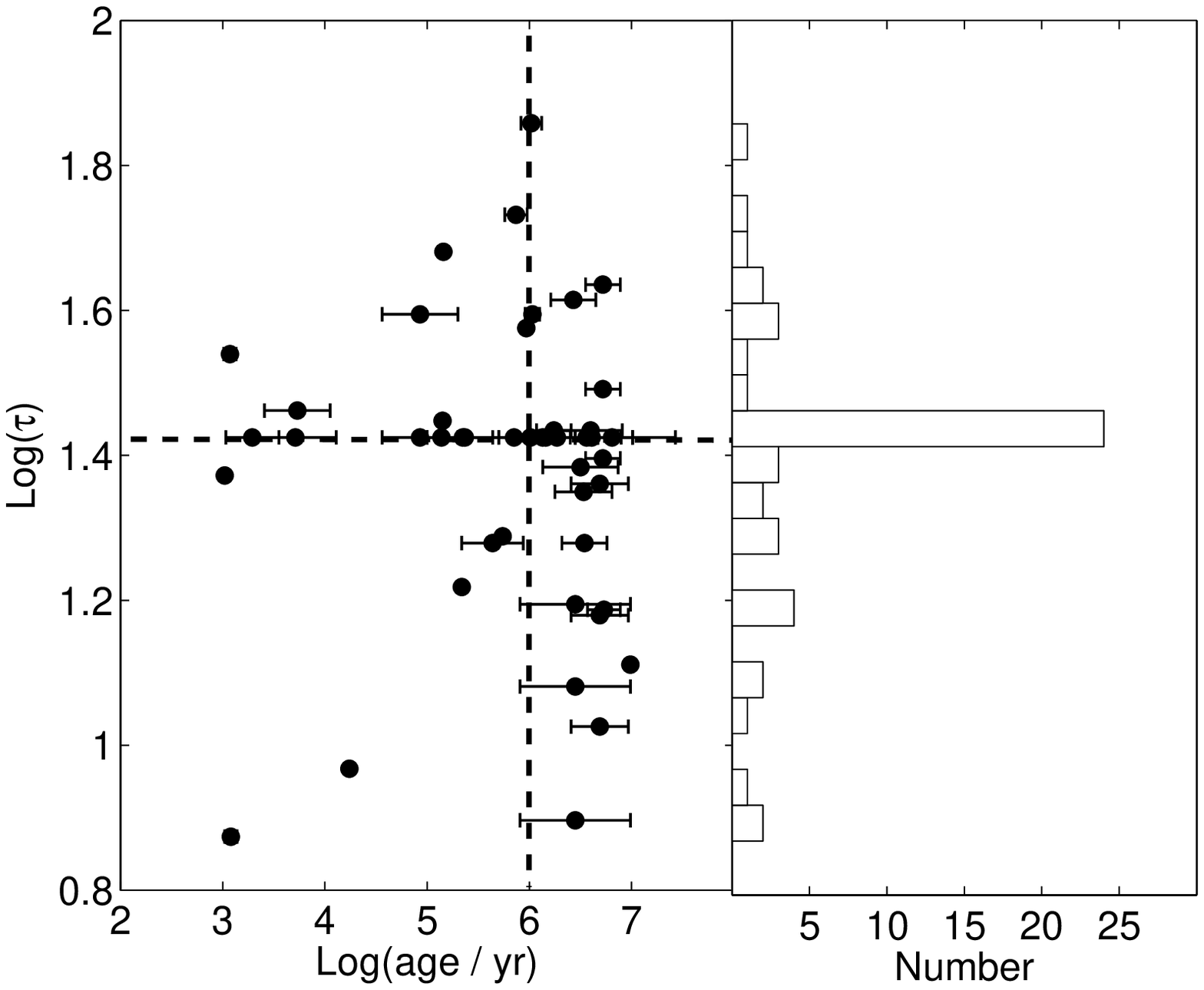}
\end{minipage}
\begin{minipage}[c]{0.5\textwidth}
  \centering
  \includegraphics[width=80mm,height=70mm,angle=0]{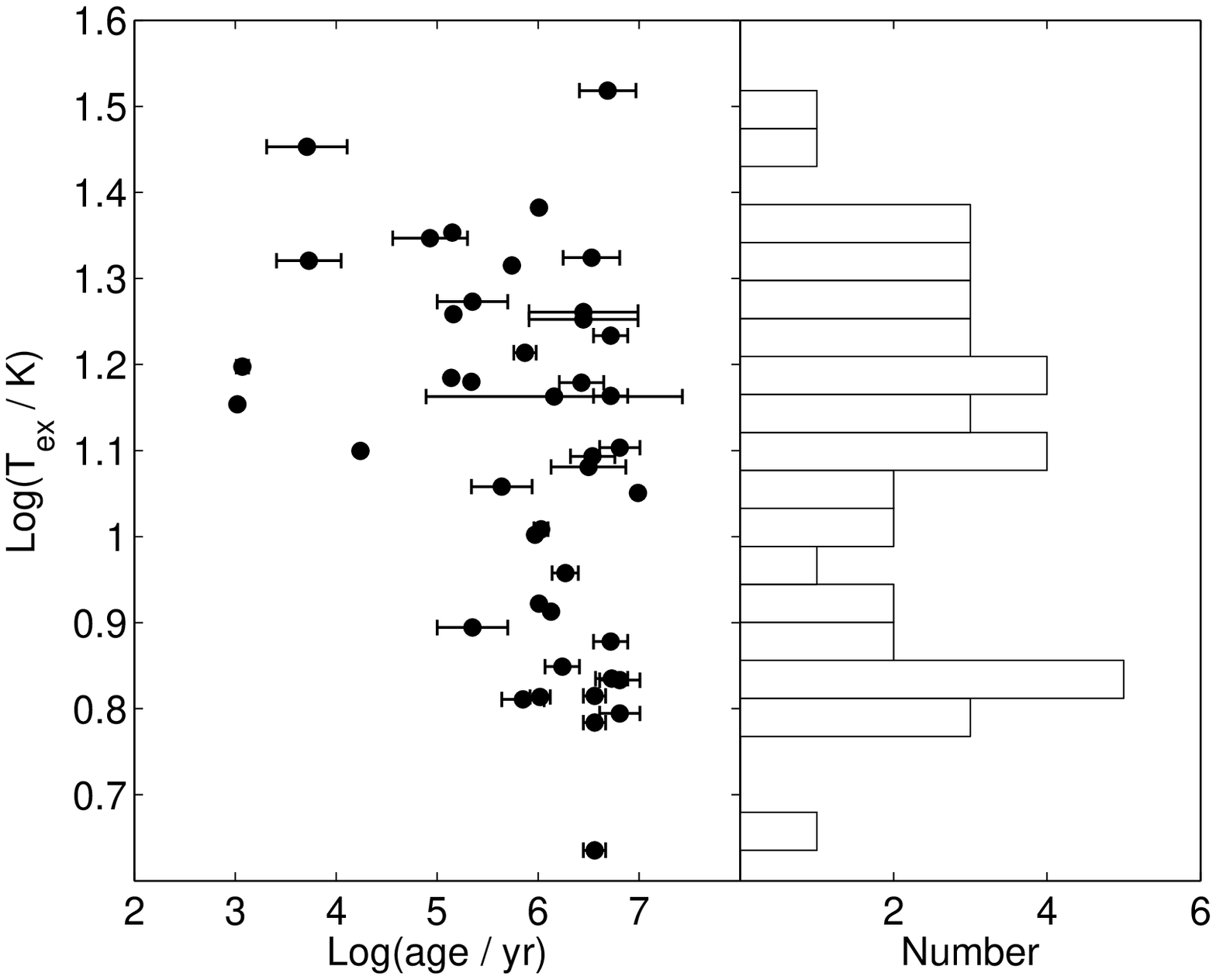}
\end{minipage}
\caption{The optical depth $\tau$ (left) of $^{12}$CO (2-1) and the
excited temperature T$_{ex}$ (right) as function of the age. The
horizonal and vertical dashed lines in the left panel mark age at
10$^{6}$ yr and the average optical depth of 26.6.}
\end{figure}

\clearpage
\begin{figure}
\begin{minipage}[c]{0.5\textwidth}
  \centering
  \includegraphics[width=70mm,height=80mm,angle=90]{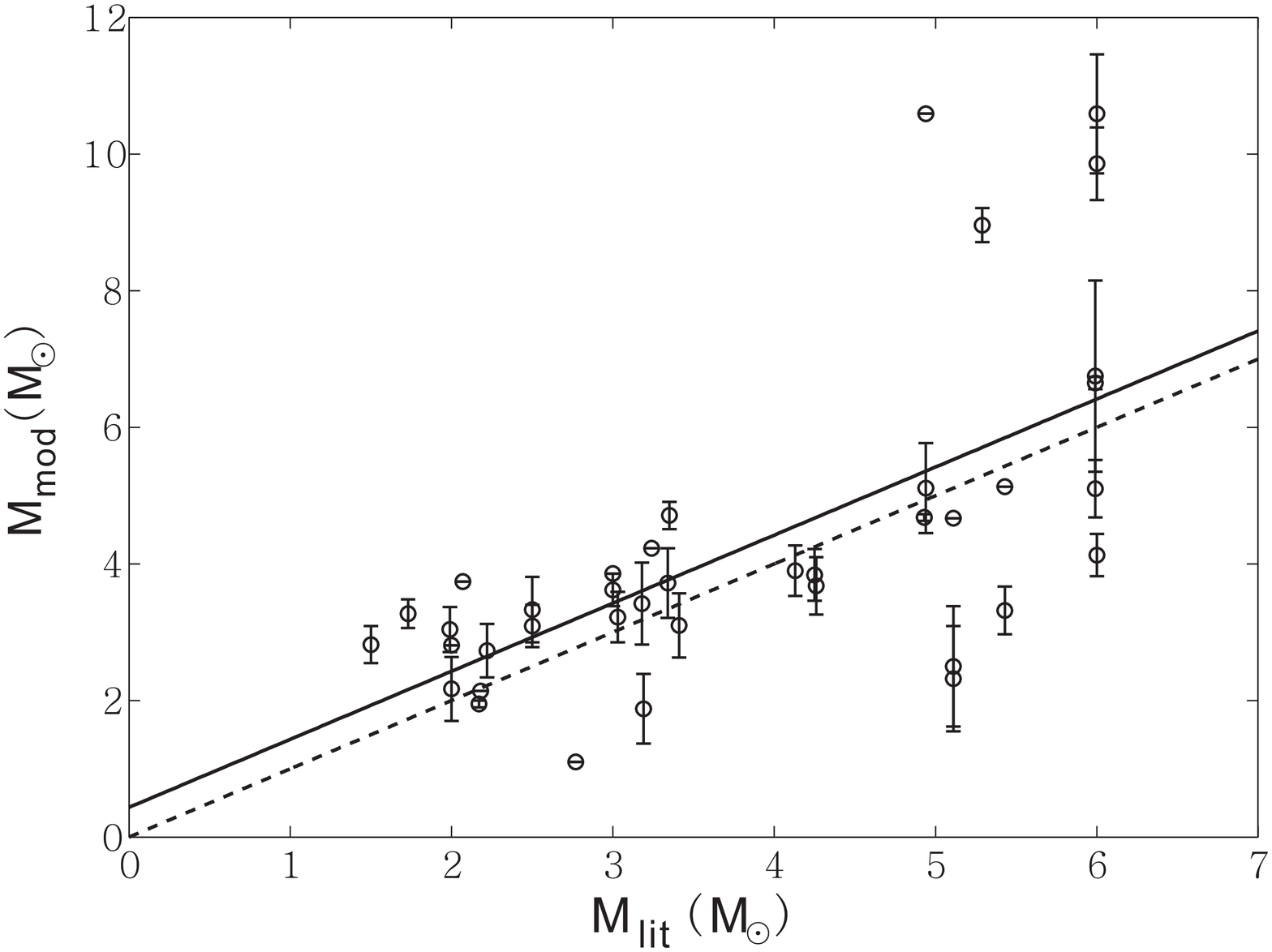}
\end{minipage}
\begin{minipage}[c]{0.5\textwidth}
  \centering
  \includegraphics[width=70mm,height=80mm,angle=90]{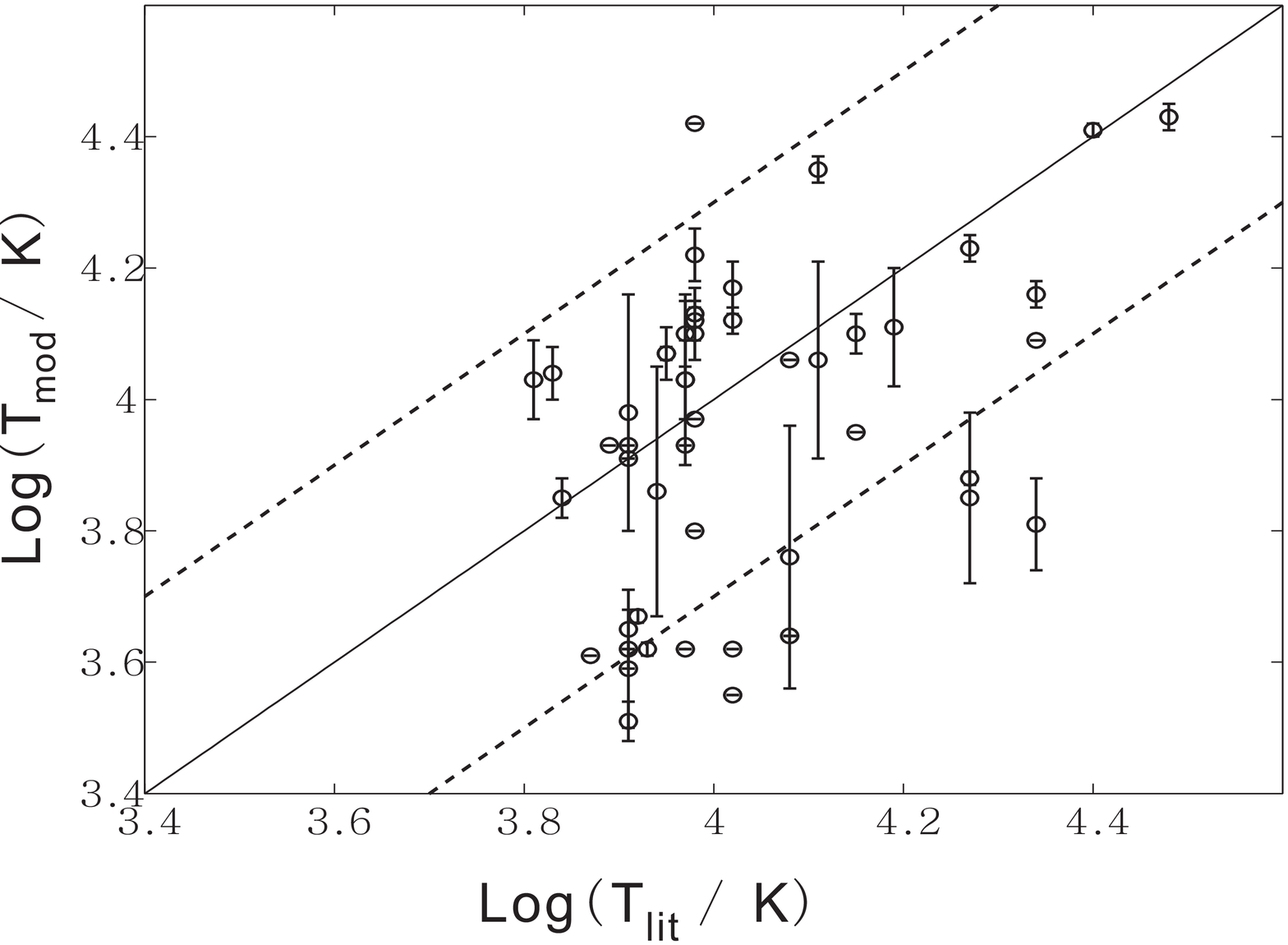}
\end{minipage}
\caption{The masses derived from SED fitting compared to those
obtained from \cite{man06} (left) and the effective temperatures of
the stars obtained from SED fitting compared to the effective
temperatures corresponding to the spectral types (right). The solid
lines in both panels describe the least-square fitting. The dashed
line in the left panel shows that M$_{mod}$ is consistent with
M$_{lit}$. The two dashed lines in the right panel mark out the area
where the effective temperatures obtained from the two independent
methods agree better than $\pm$0.3 orders of magnitude.}
\end{figure}

\begin{figure}
\includegraphics[angle=90,scale=.40]{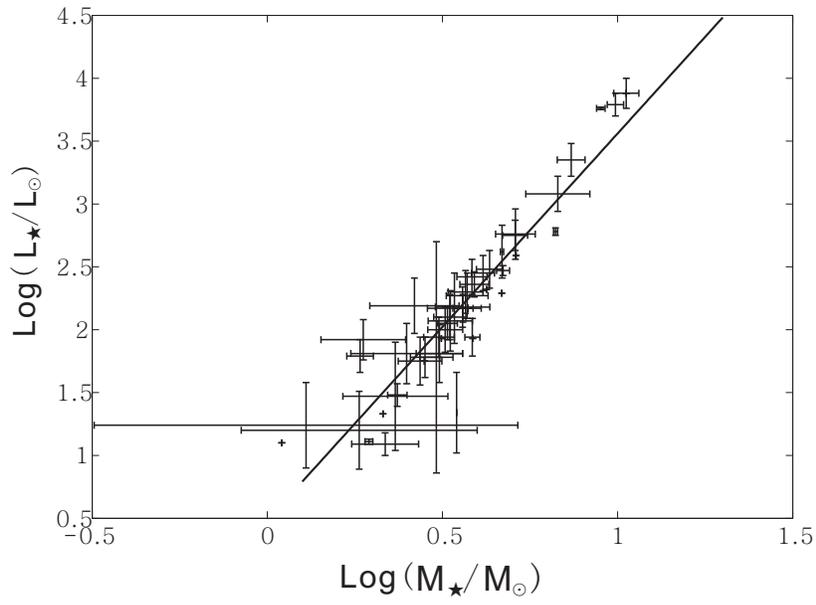}
\caption{The mass-luminosity function of the sample stars. The solid
line represents the best least-square fitting.}
\end{figure}

\clearpage

\begin{figure}
\begin{minipage}[c]{0.5\textwidth}
  \centering
  \includegraphics[width=70mm,height=80mm,angle=90]{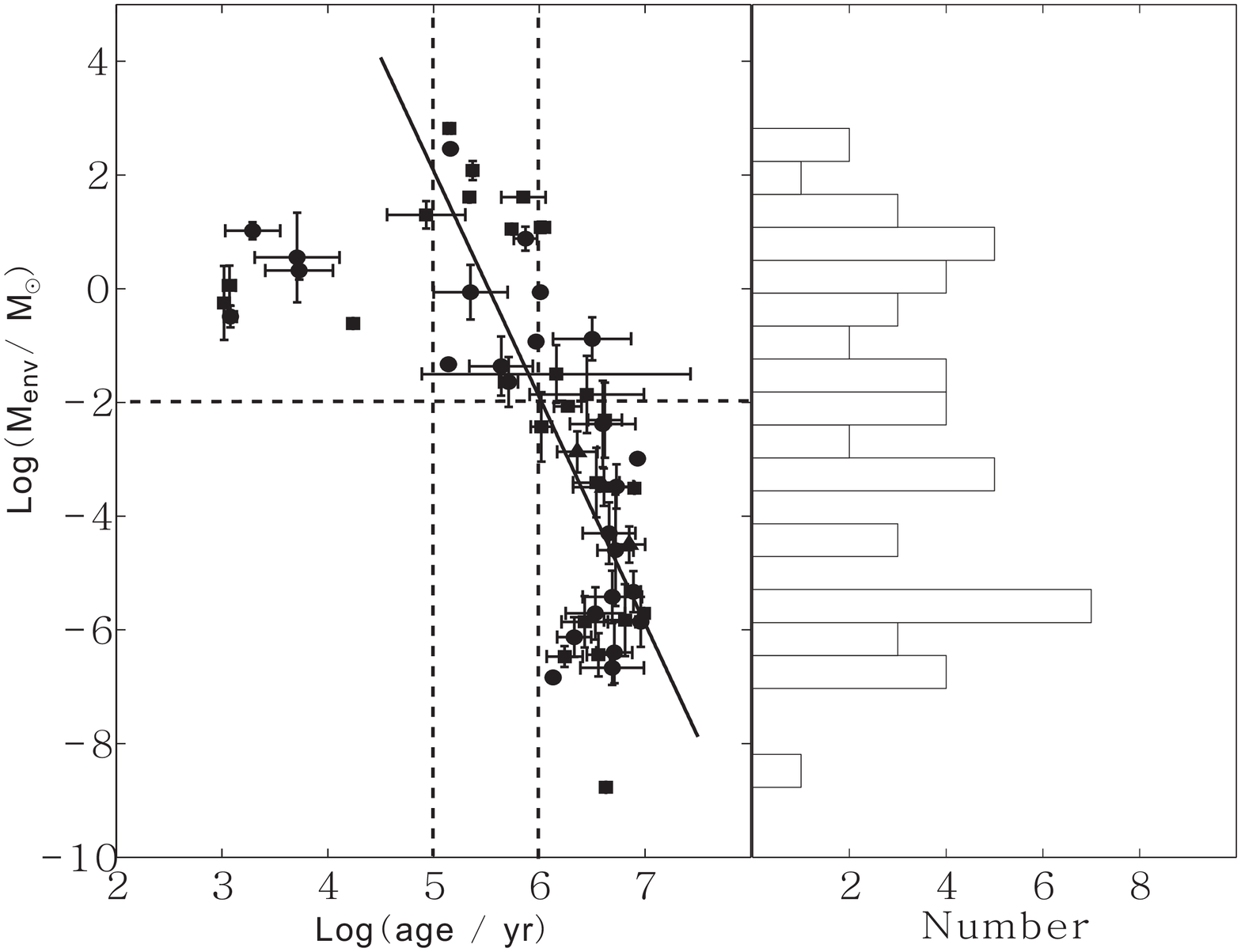}
\end{minipage}
\begin{minipage}[c]{0.5\textwidth}
  \centering
  \includegraphics[width=70mm,height=80mm,angle=90]{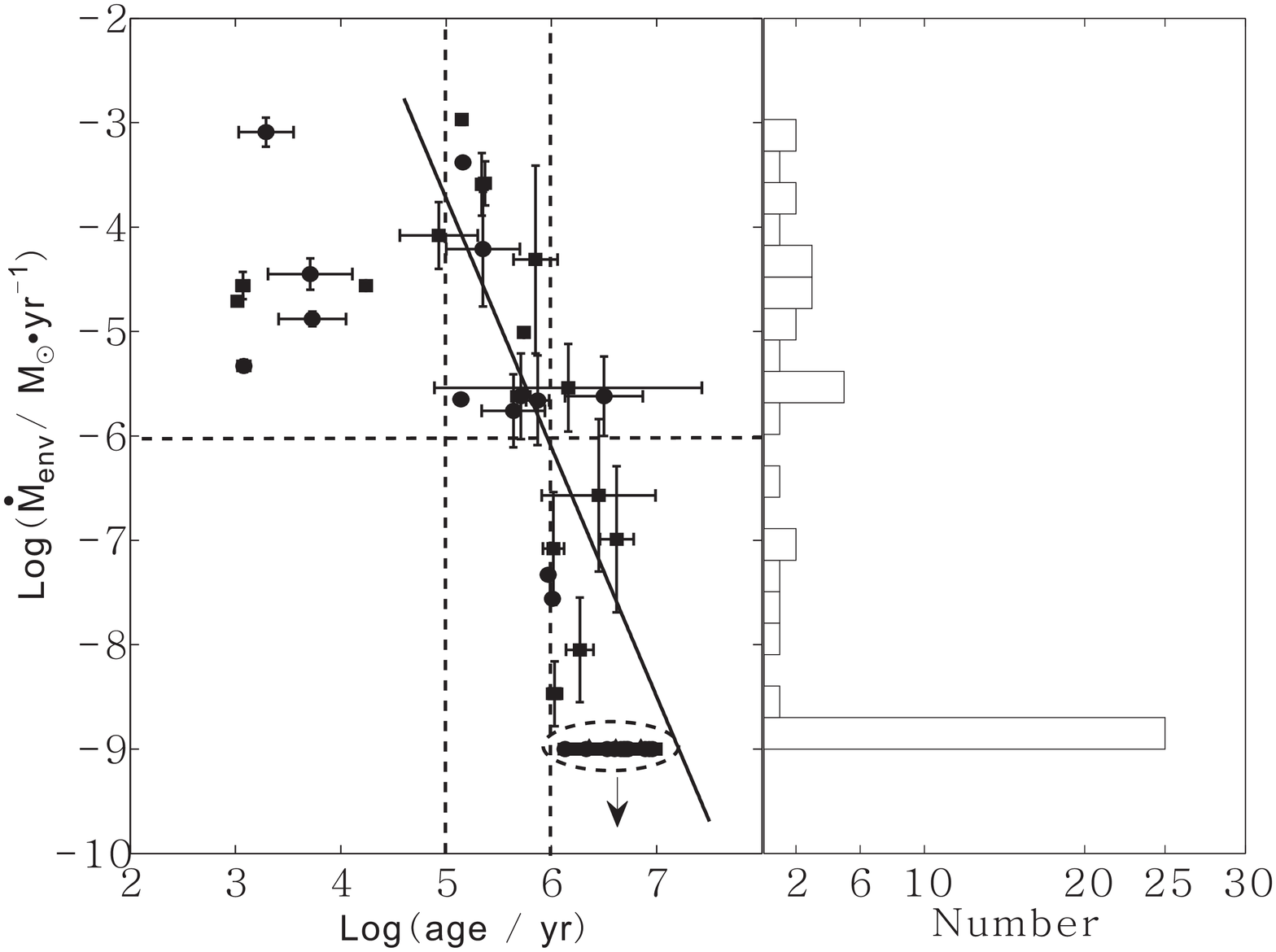}
\end{minipage}
\caption{The envelope mass (left) and the envelope accretion rate
(right) as function of the age. The solid lines represent the best
least-square fitting. The vertical dashed lines mark age at 10$^{5}$
and 10$^{6}$ yr. The horizonal dashed line in the left and right
panels mark the envelope mass of 10$^{-2}$ M$_{\sun}$ and the
envelope accretion rate of 10$^{-6}$ M$_{\sun}$yr$^{-1}$,
respectively. The dashed oval in the right panel marks the sources
with zero envelope accretion rate.}
\end{figure}

\begin{figure}
\begin{minipage}[c]{0.5\textwidth}
  \centering
  \includegraphics[width=70mm,height=80mm,angle=90]{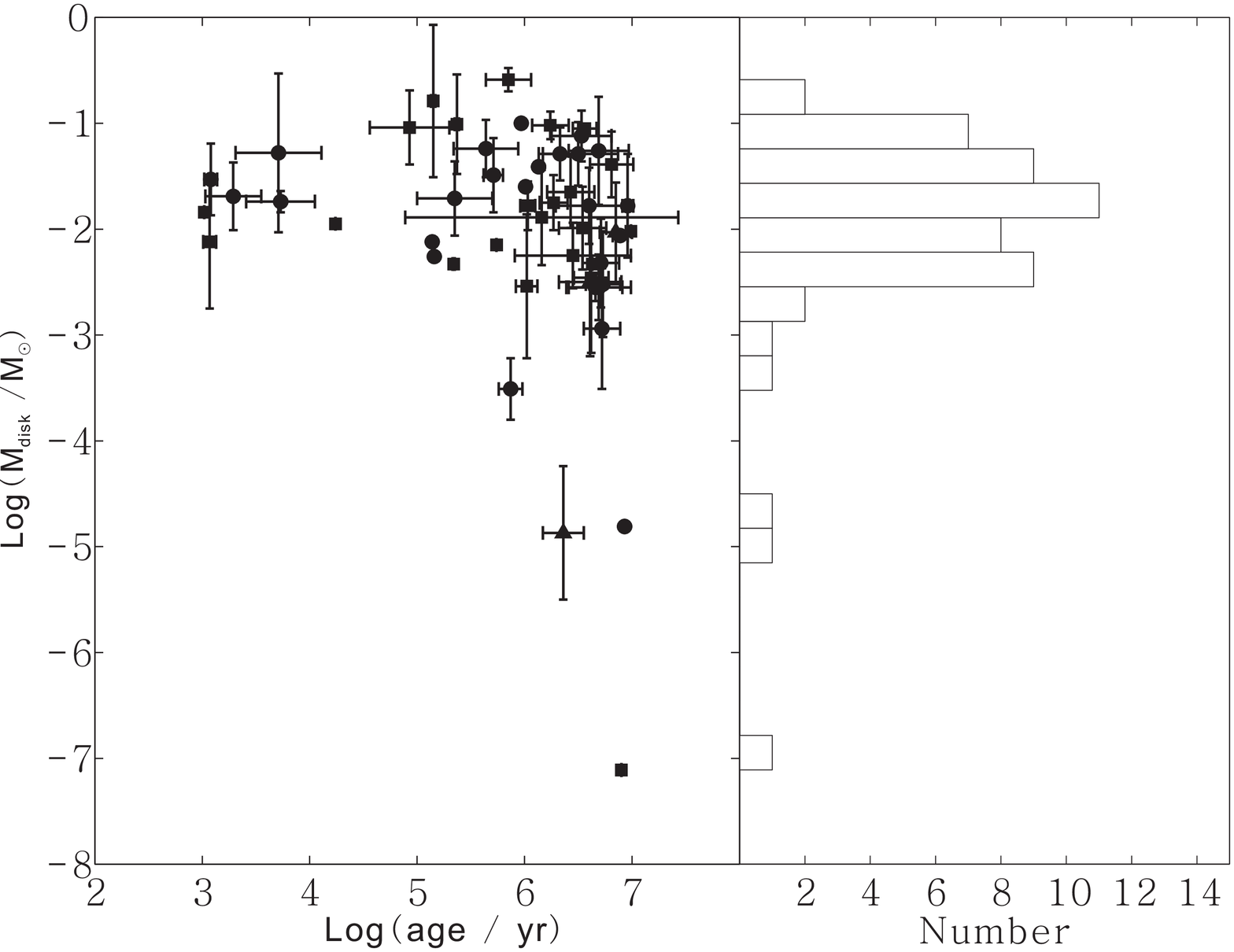}
\end{minipage}
\begin{minipage}[c]{0.5\textwidth}
  \centering
  \includegraphics[width=70mm,height=80mm,angle=90]{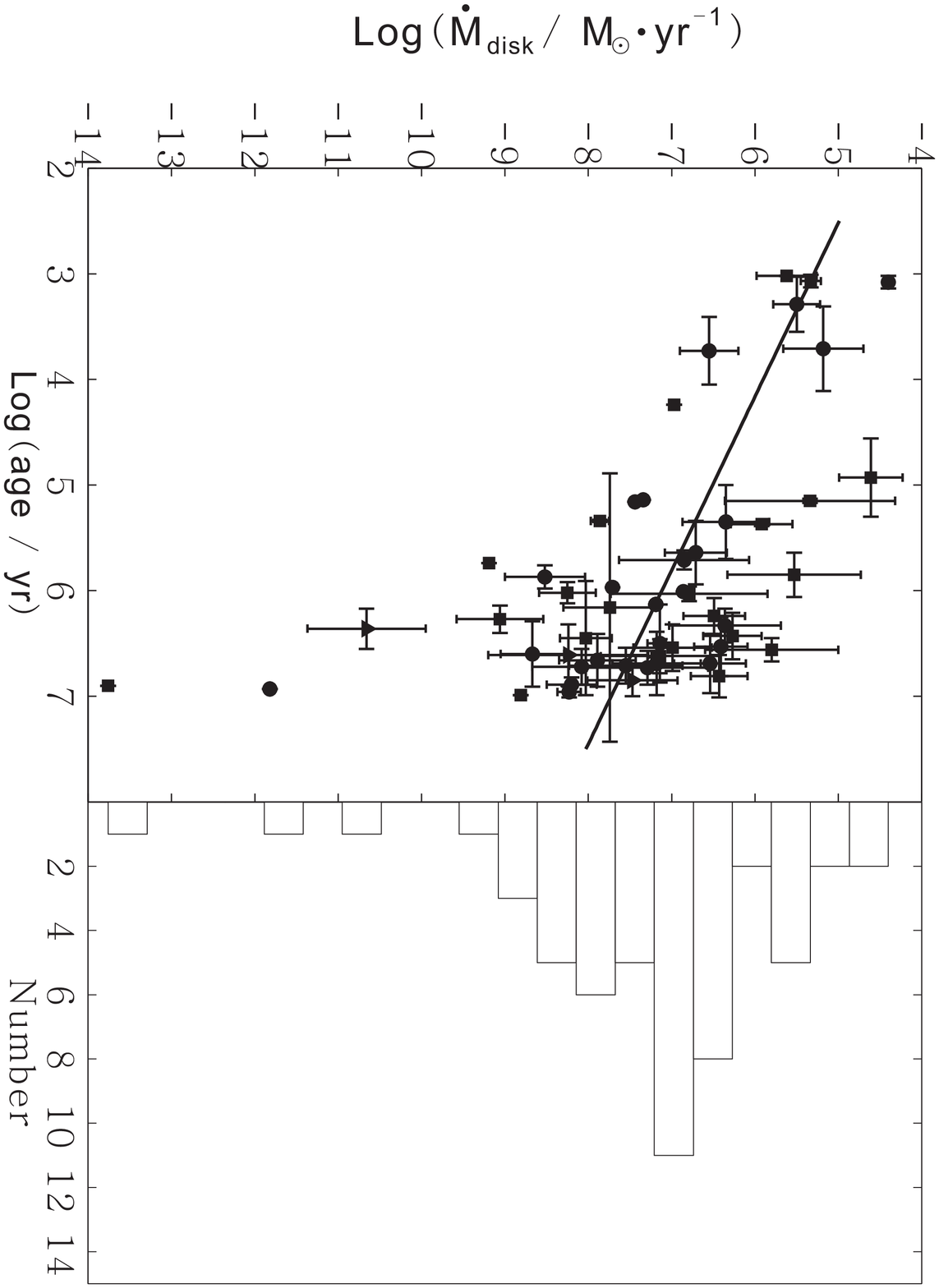}
\end{minipage}
\caption{The disk mass (left) and the disk accretion rate (right) as
function of the age. The solid lines represent the best least-square
fitting. }
\end{figure}

\begin{figure}
\begin{minipage}[c]{0.5\textwidth}
  \centering
  \includegraphics[width=70mm,height=80mm,angle=90]{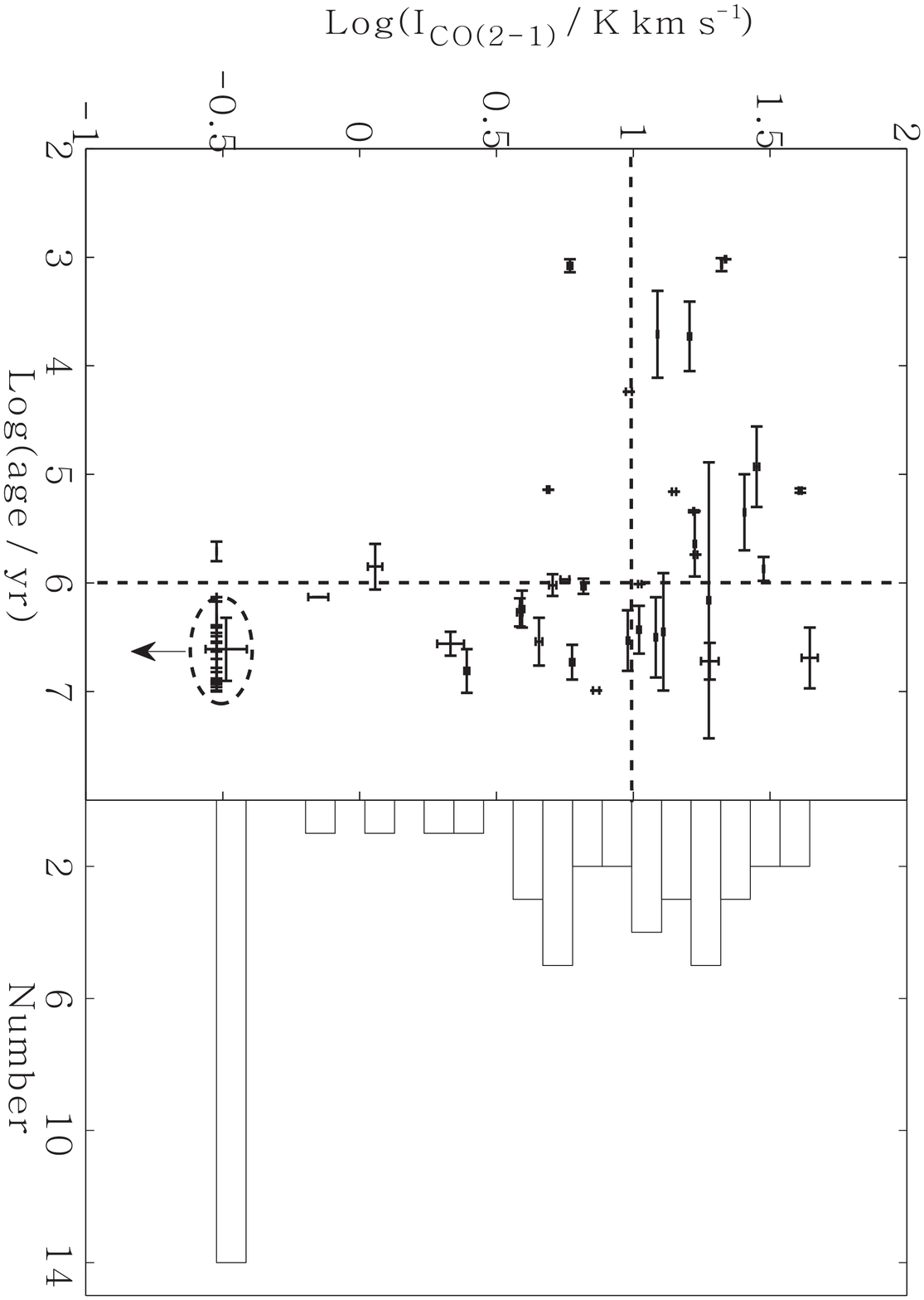}
\end{minipage}
\begin{minipage}[c]{0.5\textwidth}
  \centering
  \includegraphics[width=70mm,height=80mm,angle=90]{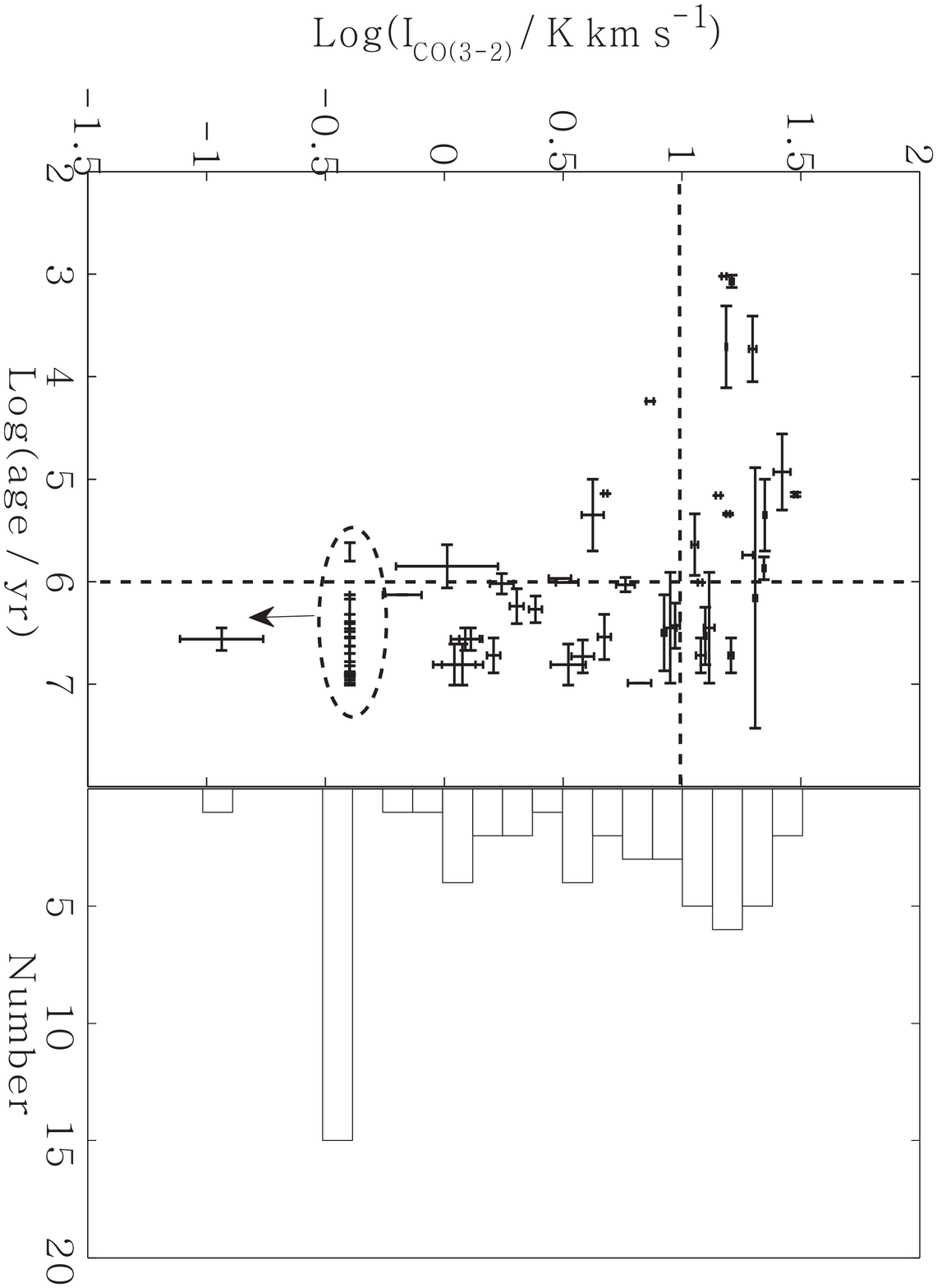}
\end{minipage}
\caption{The $^{12}$CO (2-1) intensity (left) and the $^{12}$CO
(3-2) intensity (right) as function of the age. The vertical and
horizonal dashed lines are at age equalling 10$^{6}$ yr and
$^{12}$CO intensity corresponding to 10 K~km~s$^{-1}$, respectively.
The dashed ellipses encircle the sources with low signal-to-noise
level in CO emission.}
\end{figure}

\begin{figure}
\begin{minipage}[c]{0.5\textwidth}
  \centering
  \includegraphics[width=70mm,height=80mm,angle=90]{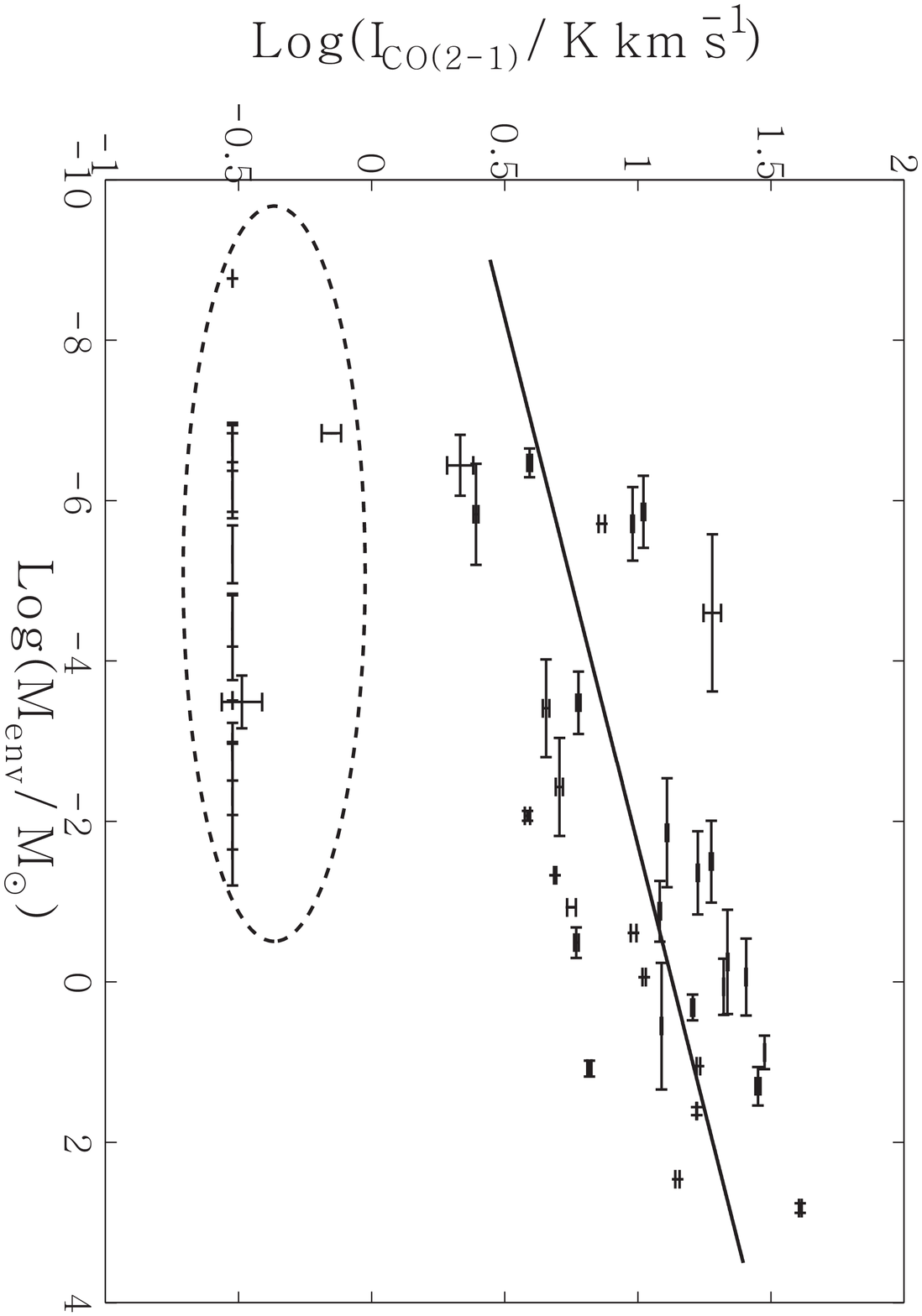}
\end{minipage}
\begin{minipage}[c]{0.5\textwidth}
  \centering
  \includegraphics[width=70mm,height=80mm,angle=90]{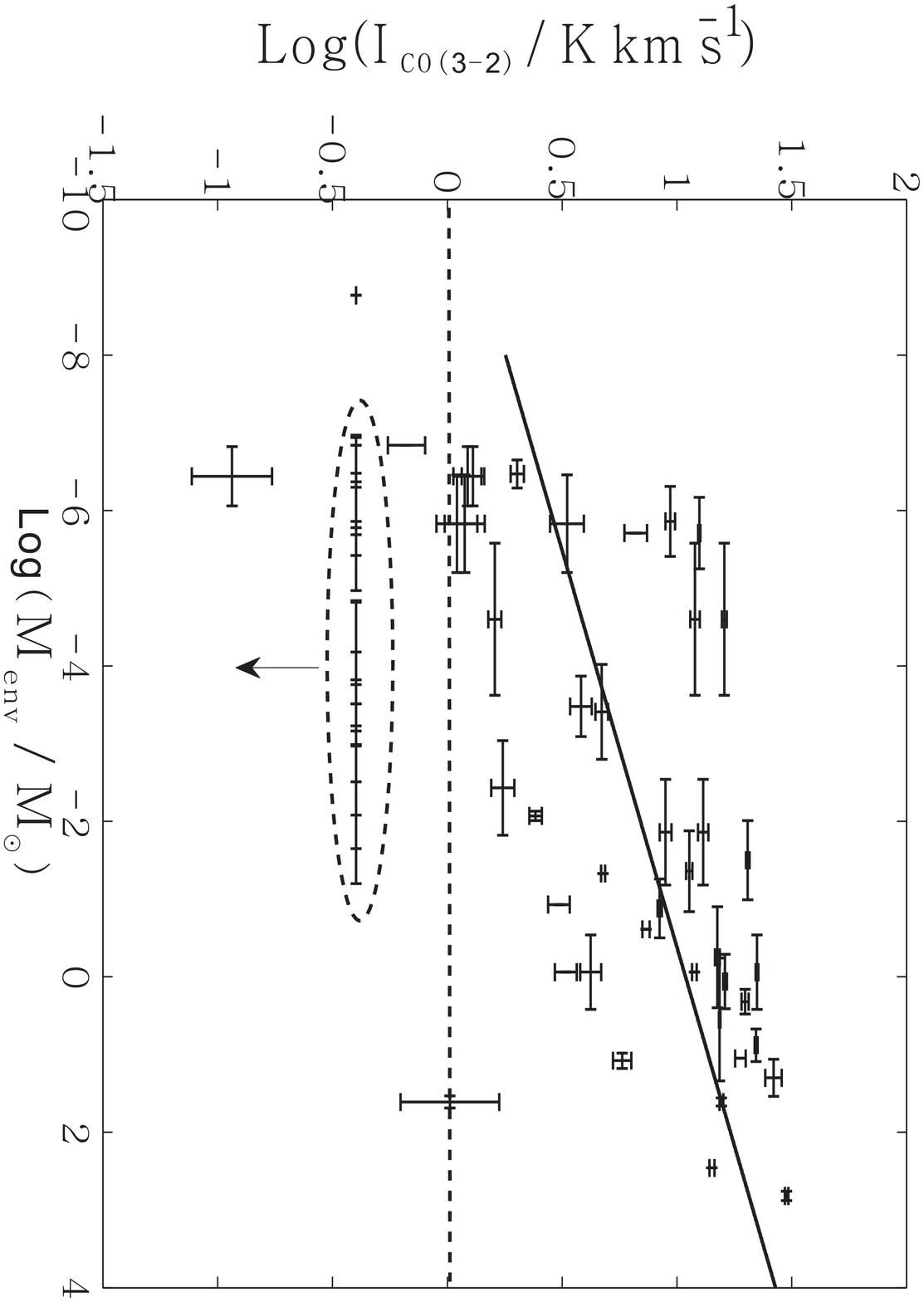}
\end{minipage}
\caption{The $^{12}$CO (2-1) intensity (left) and the $^{12}$CO
(3-2) intensity (right) as function of the envelope mass. The solid
lines represent the best least-square fitting. The horizonal dashed
line in the right panel is at where $^{12}$CO intensity corresponds
to 10 K~km~s$^{-1}$.The sources with low signal-to-noise level in CO
emission are shown in the dashed ellipses.}
\end{figure}

\end{document}